\documentclass[a4paper,onecolumn,notitlepage,nofootinbib,superscriptaddress]{revtex4-1}
\usepackage{amsmath,amssymb}
\usepackage{graphicx}
\usepackage{bm}
\usepackage{hyperref}
\usepackage{xcolor}
\usepackage{cancel}
\usepackage[normalem]{ulem}
\usepackage{aas_macros}

\newcommand{\be}{\begin{equation}}
\newcommand{\ee}{\end{equation}}
\newcommand{\derd}{\text{d}}
\newcommand{\scalp}{\!\cdot\!}
\newcommand{\la}{\left\langle}
\newcommand{\ra}{\right\rangle}
\renewcommand{\vec}{\bm}
\newcommand{\ii}{\text{i}}
\newcommand{\eh}[1]{\exp\left[#1\right]}
\newcommand{\hMpc}{h^{-1}\text{Mpc}}
\newcommand{\transp}{\text{T}}
\newcommand{\mnras}{Mon. Not. Roy. Astr. Soc.}

\begin{document}
\title{Non-perturbative halo clustering from cosmological density peaks} 
\author{Tobias Baldauf}
\email{t.baldauf@damtp.cam.ac.uk}
\affiliation{Department of Applied Mathematics and Theoretical Physics, University of Cambridge, CB3 0WA, Cambridge, UK}
\author{Sandrine Codis}
\affiliation{CNRS \& Sorbonne Universit\'e, UMR 7095, Institut d'Astrophysique de Paris, 75014, Paris, France}
\affiliation{IPhT, DRF-INP, UMR 3680, CEA, Orme des Merisiers Bat 774, 91191 Gif-sur-Yvette, France}
\author{Vincent Desjacques}
\affiliation{Physics Department and Asher Space Science Institute, Technion, Haifa 3200003, Israel}
\author{Christophe Pichon}
\affiliation{CNRS \& Sorbonne Universit\'e, UMR 7095, Institut d'Astrophysique de Paris, 75014, Paris, France}
\affiliation{IPhT, DRF-INP, UMR 3680, CEA, Orme des Merisiers Bat 774, 91191 Gif-sur-Yvette, France}
\affiliation{KIAS, 85 Hoegiro, Dongdaemun-gu, 02455 Seoul, Republic of Korea}
\begin{abstract}
Associating the formation sites of haloes with the maxima of the smoothed linear density field, we present non-perturbative predictions for the Lagrangian and evolved halo correlation functions that are valid at all separations. In Lagrangian space, we find significant deviations from the perturbative bias calculation at small scales, in particular, a pronounced exclusion region where $\xi=-1$ for maxima of unequal height. Our predictions are in good agreement with the Lagrangian clustering of dark matter proto-haloes reconstructed from $N$-body simulations. Our predictions for the mean infall and velocity dispersion of haloes, which differ from the local bias expansion, show a similar level of agreement with simulations. Finally, we displace the initial density peaks according to the Zeldovich approximation in order to predict the late-time clustering of dark matter haloes. While we are able to reproduce the early evolution of this conserved set of tracers, our approximation fails at the collapse epoch ($z=0$) on non-linear scales $r\lesssim 10\hMpc$, emphasizing the need for a non-perturbative treatment of the halo displacement field.
\end{abstract}

\maketitle

\section{Introduction}

Upcoming galaxy surveys will map out the positions of galaxies in the Universe over large volumes with unprecedented precision \cite{Hill:2008mv,Euclid,Green:2012mj,Alam:2016hwk,Abbott:2017wau}. A detailed understanding of the clustering statistics of these luminous tracers of the large-scale matter distribution is necessary in order to harvest the late time Universe and use it as a laboratory for fundamental physics.
However, as recognized long ago, galaxy formation in Cold Dark Matter (CDM) cosmologies preferentially takes place inside the potential wells of virialized CDM structures or haloes \cite{binney:1977,rees/ostriker:1977,white/rees:1978}. This allows us to consider the somewhat simpler problem of halo clustering, which can be expressed as a biased version \cite{kaiser:1984,bbks,szalay:1988,cole/kaiser:1989,mo/white:1996,sheth/tormen:1999} of the clustering of the underlying matter distribution (see \cite{biasreview} for a recent review).
Simulations provide a straightforward way to investigate halo clustering and allow us to study the phenomenology of their $n$-point functions. Notwithstanding, the theoretical understanding of the measured correlation functions is still rudimentary for all but the largest scales. 

The clustering of peaks of the 3-dimensional density field as a proxy for the initial formation sites of virialized structures was pioneered by \citep{bbks}, following earlier works by \citep{gunn/gott:1972} on the collapse of spherical overdensities and by \citep{press/schechter:1974} on a statistical approach to the distribution of virialized structures. Peak correlations were initially investigated for Gaussian random fields in the high-threshold limit in order to explain the bias of massive clusters  \citep{kaiser:1984,cline/etal:1987}.
Later, these calculations were extended to include exclusion \cite{Appel:1990mfb}, arbitrary peak thresholds \citep{matsubara:1995,regos/szalay:1995}, non-Gaussianities \citep{2012PhRvD..85b3011G} and anisotropies to incorporate redshift space distortions \citep{desjacques/sheth:2010,2013MNRAS.435..531C,Matsubara:2019tyb}. 
With the identification of virialized cosmological objects with a point process of ``peak patches'' in the initial (Lagrangian) space \cite{Bond:1993wd,Bond:1993we,Stein:2018lrh} (see also \cite{musso/sheth:2019}), the peak approach provided a useful framework to understand the scale-dependence, stochasticity and gravitational evolution of the clustering of virialized structures \citep{Desjacques:2008jj,Desjacques:2010gz,Baldauf:2016aaw}. While the association between virialized dark matter haloes and peaks of the initial density field is fairly tight for objects of mass $M\gg M_*$, it weakens with decreasing $M$, as shown by detailed studies of cosmological numerical simulations \citep{Ludlow:2011tpf,anna/ludlow/porciani:2012,hahn/paranjape:2014}. However, an identification based on a local energy definition can improve the correspondence \cite{musso/sheth:2019}.
Overall, this approach can be extended to include (local and non-local) constraints motivated by non-linear structure formation (along the lines of, e.g. \cite{vdw/bertschinger:1996,paranjape/sheth:2012,castorina/paranjape/etal:2017}), or embedded in an effective field theory \citep{vlah/castorina/white:2016}. Nonetheless, a simpler Lagrangian peak constraint a la \cite{bbks} (possibly combined with excursion set theory, see \cite{paranjape/sheth:2012}) already provides a useful framework to understand the bias arising in clustering statistics of non-linear Large-Scale Structure (LSS). For instance, it provides a physical explanation for the assembly bias of massive haloes \cite{dalal/white/etal:2008,2018MNRAS.476.4877M}, and their sub-Poissonian noise \cite{Baldauf:2013hka}. It also predicts the low-order halo bias parameters (from a two-parameters only description of the halo collapse barrier) with reasonable accuracy \citep{paranjape/etal:2013,Biagetti:2013hfa,lazeyras/etal:2016,Lazeyras:2019dcx} (see also \cite{chirag/castorina/seljak:2017} for an extension to the tidal shear bias).
Furthermore, it can be implemented to investigate the merger history of haloes (and other cosmic web components) across cosmic time \citep{2020MNRAS.496.4787C}.

{The Baryon Acoustic Oscillation (BAO) feature in the galaxy correlation function and power spectrum has proven to be an important ruler for measuring the expansion history of the Universe and inferring the equation of state of dark energy. Large scale motions degrade the linear BAO feature, motivating so called reconstruction methods \cite{Eisenstein:2006nk,Noh:2009bb,Padmanabhan:2008dd} (see also \cite{peebles:1989,bertschinger/etal:1990,nusser/dekel:1992,nusser/branchini:2000} for early work on the topic) that aim to undo the effect of the bulk motions. Most of the studies of BAO smoothing and reconstruction are based on the local bias model, where the BAO in Lagrangian space is given by linear theory and halo motions are unbiased with respect to the dark matter displacement. It has been shown that peaks show a more pronounced BAO feature, in agreement with what is observed for proto-haloes in Lagrangian space \cite{Baldauf:2016aaw}. Furthermore, halo velocities do deviate from the underlying dark matter velocities on all but the largest scales, again in agreement with what is predicted by the peak model \cite{Achitouv:2015gma,Baldauf:2016aaw}. In this study we will present a detailed comparison of mean-streaming and displacement dispersion measurements of peaks and haloes in simulations.}

Peak theory in a broad sense has also been extremely successfully over the past decades in describing the cosmic web formation and evolution, including clusters but also filaments, walls and voids. 
Building on the seminal work of \cite{Bond:1995yt}, the skeleton picture \citep{2009MNRAS.393..457S} extended peak theory and was able to accurately describe the fully connected cosmic-web, its length and curvature \citep{2009MNRAS.396..635P}, its connectivity \citep{2018MNRAS.479..973C}, and its impact on galaxy formation \citep{2018MNRAS.474..547K,2015MNRAS.452.3369C}. Specific works also focused on voids \citep[e.g]{2013MNRAS.434.2167J,2018MNRAS.475.1912A} or saddle points only \citep{2019MNRAS.489..900F} and their respective clustering properties \citep{2020arXiv201104321S}, and cosmic web configurations in the initial conditions \cite{aung/cohn:2016}.

In this paper, we extend our previous work on the non-perturbative peak correlation function in one spatial dimension \cite{Baldauf:2015fbu} (hereafter BCDP) to the more realistic 3-dimensional (3D) case.
First, in Section~\ref{sec:Lag}, the formalism to predict the clustering of peaks in the initial Gaussian density field is described and explicit correlations in the simpler case of signed critical points are derived.
Section~\ref{sec:MCMC} then computes numerically the peak correlations and compares to a large-scale bias expansion for fixed peak heights and bins. The induced shot noise correction is discussed, while pairwise velocity statistics in the peak model are computed and compared again to first-order bias expansions.
From the statistical knowledge of the velocity field, we then study the time evolution of peak clustering by taking into account their Zeldovich displacement in Section~\ref{sec:evol}. Finally, Section~\ref{sec:realizations} compares our results to the statistics of peaks in random field realizations. We wrap up in Section~\ref{sec:conclusion}.
When making predictions for realistic $\Lambda$CDM cosmologies throughout this paper, we will consider a WMAP7 cosmology with parameters $\Omega_\text{m}=0.272$, $\sigma_8=0.81$, $n_\text{s}=0.967$. We will also consider power law power spectra of the form $P_\text{lin}=A k^{n_\text{s}}$.

\section{Peak Clustering in Lagrangian Space}
\label{sec:Lag}

The statistical properties of the Gaussian field and its derivatives are fully encoded in the multi-point moments of its power spectrum. The variance of the field and its derivatives is given by
\be
\sigma_{i}^2=\int \frac{\derd^3 k}{(2\pi)^3} P_\text{s}(k) k^{2i}\; ,
\ee
where $P_\text{s}(k)=P_\text{lin}(k)W^2_R(k)$ is the filtered linear matter power spectrum and the filter is taken to be a Gaussian for definiteness, $W_R(k)=\exp(-k^2 R^2/2)$. Note however that the effective halo window function has been shown to be different from a Gaussian, see for instance \cite{chan/sheth/scoccimarro:2017}.
It is convenient to introduce the spectral parameters $\gamma\equiv\sigma_1^2/\sigma_0/\sigma_2$ and $R_\star\equiv\sigma_1/\sigma_2$ which quantify the width of $P_\text{s}(k)$ and the characteristic radius of the peaks, respectively. For later convenience, we will also define the velocity bias as $R_v \equiv \sigma_0/\sigma_1$.
The correlation between the field properties at distinct locations are given by the correlation functions
\be
\xi_{i,l}(r)=\int \frac{\derd^3 k}{(2\pi)^3} P_\text{s}(k) k^i j_l(k r)\; ,
\ee
where $j_l$ are the spherical Bessel functions of order $l$.

The number density of maxima in a smoothed field $\delta_\text{s}$ at Lagrangian position $\vec q$ is given by the set of points that have a vanishing gradient and negative definite Hessian
\be
n(\vec q)=|\det H(\vec q)|\delta^\text{(D)}(\sigma_1 \vec \eta)\Theta(-\max_i \lambda_i)\, ,
\label{eq:peakabun}
\ee
where $\sigma_2 \lambda_i$ are the ordered ($\lambda_3<\lambda_2<\lambda_1$) eigenvalues of the Hessian $H_{ij}=\partial_i\partial_j \delta_\text{s}$
and $\sigma_1 \vec \eta=\vec \nabla \delta_\text{s}$ is the gradient of the field. 
The symmetric Hessian matrix $H_{ij}$ has six independent components 
\be
\sigma_2 \zeta_1=H_{11},\sigma_2\zeta_2=H_{22},\sigma_2\zeta_3=H_{33},\sigma_2\zeta_4=H_{12},\sigma_2\zeta_5=H_{13},\sigma_2\zeta_6=H_{23}\,,
\ee
such that the determinant yields
\be
\det H=\sigma_2^3\left(\zeta_1 \zeta_2 \zeta_3+2\zeta_4\zeta_5\zeta_6-\zeta_1 \zeta_6^2 -\zeta_3\zeta_4^2-\zeta_2\zeta_5^2\right)=\sigma_2^3 \lambda_1 \lambda_2\lambda_3 \, .
\ee
Furthermore, the trace reads ${\rm tr} H = \sigma_2\zeta$, where the peak curvature $\zeta$ is given by
\be
\zeta=\zeta_1+\zeta_2+\zeta_3=\sum_i\lambda_i \, .
\ee
Together with the gradient of the field and the field itself, we thus have to consider ten field variables at each of the points under consideration. This number would increase to thirteen should we also consider peak velocities or displacements. Based on the spherical collapse model, we expect that overdense perturbations collapse into haloes whenever they cross a critical collapse threshold $\delta_\text{c}$ on a given smoothing scale. It is often convenient to express the overdensity at the peak location in terms of the peak height or significance $\nu=\delta/\sigma_0$.

The mean abundance of peaks is given by \citep{Bardeen:1985tr}
\be
\bar n=\langle n(\vec q)\rangle=\int \derd \vec X w(\vec X) \mathbb{P}_\text{1pt}(\vec X),
\ee
where $\mathbb{P}_\text{1pt}$ is the one-point distribution of the field and its first and second derivatives gathered in the state vector $\vec X^\text{T}=(\sigma_0 \nu, \sigma_1 \vec \eta, \sigma_2 \vec \zeta)$ and $w(\vec X)$ is the localised peak number density {\it in field space} encoding the peak weight 
\be
w(\vec X)=|\det H|\delta^\text{(D)}(\sigma_1 \vec \eta)\Theta(-\max_i \lambda_i).
\label{eq:weight}
\ee
We will often impose further constraints on the peak height and consider fixed peak heights encoded by a Dirac delta function or bins in peak height encoded by a top-hat window.

To study the clustering of peaks, let us introduce now 
their two-point correlation function as
\be
1+\xi(r)=\frac{1}{\bar n^2}\int \derd \vec X_1 \int \derd \vec X_2 w(\vec X_1)w(\vec X_2) \mathbb{P}_\text{2pt}(\vec X_1,\vec X_2|r),
\ee
where the two-point probability distribution function is given as a multivariate Gaussian
\be
\mathbb P_\text{2pt}(\vec X_1,\vec X_2)=\frac{1}{\sqrt{(2\pi)^{20} |\det \vec C|}}\exp\left[-\frac{1}{2}\vec X^{\rm T} \scalp \vec C^{-1} \scalp\vec X \right] ,
\ee
of the joint state vector $\vec X^\text{T}=(\vec X_1^\text{T},\vec X_2^\text{T})$ at the two spatial positions $\vec r_1$ and $\vec r_2$ with $r=|\vec r_2-\vec r_1|$. The covariance matrix $C_{ij}=\la X_{i} X_{j}\ra$ can be explicitly computed from the power spectrum \citep{Bardeen:1985tr}.

As explained in BCDP and \cite{Matsubara:2019sab}, the peak correlation can be evaluated by drawing samples from the conditional distribution of peak curvatures given the peak height and a vanishing density gradient. 
Reordering and splitting the state vector as $\vec X^\text{T}=(\vec m^\text{T},\vec n^\text{T})$ with $\vec m^\text{T}=(\sigma_0 \nu_1,\sigma_1 \vec \eta_1,\sigma_0 \nu_2,\sigma_1 \vec \eta_2)$ and $\vec n^\text{T}=(\sigma_2 \vec \zeta_1,\sigma_2 \vec \zeta_2)$, we can write down the conditional probability of peak curvatures given the peak amplitude and vanishing gradient $\mathbb P(\vec m \cap \vec n)=\mathbb P(\vec m)\mathbb P(\vec n|\vec m)$.
Starting from a Cholesky decomposition of the covariance matrix of the curvature component\footnote{We will split the covariance matrix and its inverse as\begin{align}
\vec C=\begin{pmatrix}
\vec C_{\vec m,\vec m} & \vec C_{\vec m,\vec n}\\
\vec C_{\vec m,\vec n}^\transp & \vec C_{\vec n,\vec n}
\end{pmatrix}\, ,
&&
\vec \Omega=\vec C^{-1}=\begin{pmatrix}
\vec \Omega_{\vec m,\vec m} & \vec \Omega_{\vec m,\vec n}\\
\vec \Omega_{\vec m,\vec n}^\transp & \vec \Omega_{\vec n,\vec n}
\end{pmatrix}\, .
\end{align}}
$\vec \Omega_{\vec n,\vec n}^{-1}=\vec Q \vec Q^\text{T}$ and a vector $\vec N_0$ of normal distributed random numbers, we can generate a sample of the conditional distribution of peak curvatures as
\be
\vec N=\vec \mu_{\vec n} + \vec Q^\text{T} \vec N_0\; ,
\ee
where $\vec \mu_{\vec n}=-\vec m^{\rm T} \vec C_{\vec m,\vec m}^{-1}\vec C_{\vec m,\vec n}$.
To efficiently check the negative definiteness of the Hessian, we use the Sylvester criterion, first checking $H_{11}<0$ then $H_{11}H_{22}-H_{12}^2>0$ and finally $\det H<0$ for both points. If all of these criteria are satisfied we add up the absolute values of the determinant in Eq.~\eqref{eq:weight}.
\subsection{Signed Critical Points}\label{sec:signedcritpoints}
As we have pointed out above, the evaluation of the peak correlation function requires a numerical sampling of the components of the Hessian at the two locations. We thus cannot write down a closed form analytic expression for the peak correlation function. What prevents us from doing so is the absolute value of the determinant in Eq.~\eqref{eq:weight} and the negative definiteness constraint. Without these two complications, we can indeed derive a closed form expression for signed critical points. A similar calculation was performed in \cite{Verde:2014nwa}, where the determinant weight in Eq.~\eqref{eq:peakabun} was dropped altogether, by weighting with $1/|\det H|$.

The expected abundance of signed critical points with height or significance, $\nu$, is given by
\be
\bar n_\text{crit}=\frac{e^{-\frac{\nu ^2}{2}} \gamma^3 \nu  \left(\nu ^2-3\right) }{12 \sqrt{3} \pi ^2 R_\star^3}\,.
\ee
This formula can be obtained along the lines of the BBKS derivation of the peak abundance, or upon integrating out the curvature variables after rewriting the prefactor as derivatives. 
For a large significance $\nu\gg 1$, this abundance agrees with the abundance of maxima, since all high extrema are maxima. Note also that it is equivalent to the derivative of the Euler characteristic.

\begin{figure}
\centering
\includegraphics[width=0.5\textwidth]{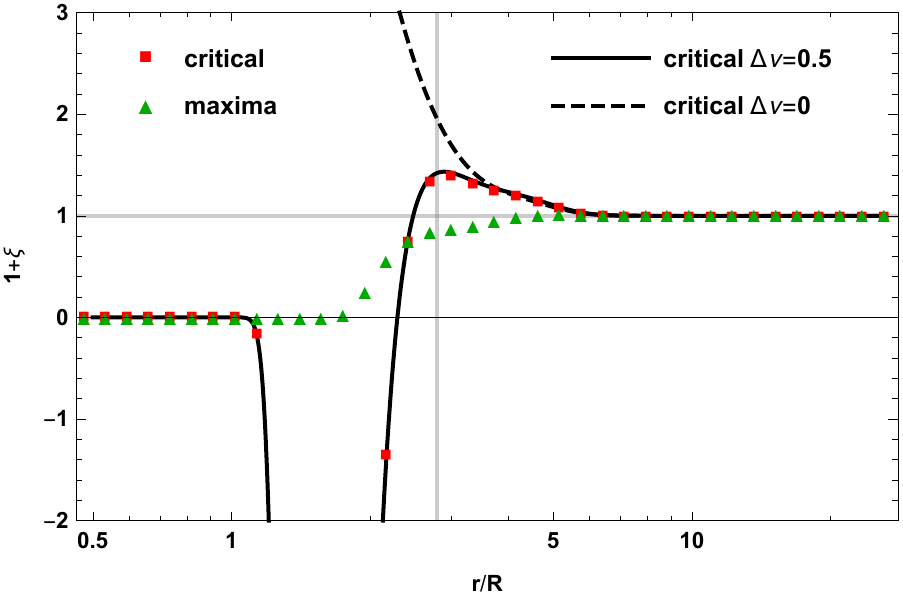}
\caption{Correlation function of signed critical points or extrema in a density field characterized by a $n_\text{s}=0$ power law power spectrum. We compare our analytical formula for $\Delta \nu=0$ (dashed) and $\Delta \nu=0.5$ (solid) with a numerical implementation of critical points (red squares) and peaks (green triangles) in the $\Delta \nu=0.5$ case. The vertical line indicates the estimate of the exclusion scale as given by Eq.~\eqref{eq:exclscale}.}
\label{fig:critpoints}
\end{figure}

The full result for the correlation function of signed critical points is derived in Appendix~\ref{app:critpoints} and has the form
\be
1+\xi(r)=A(r)e^{B(r)}\, .
\ee
For a power-law power spectrum with $n=0$ and for critical points with fixed heights $\nu+\Delta\nu/2$ and $\nu-\Delta\nu/2$, the exponential is given by 
\be
B(r)=-\frac12 {\vec m}^{\rm T} C_{\vec m,\vec m}^{-1} {\vec m}=\frac{\nu ^2}{4}+\frac{7 \Delta \nu ^2}{80}-\frac{96 \Delta \nu
   ^2}{\tilde r^6}+\frac{12 \Delta \nu
   ^2}{\tilde r^4}-\frac{6 \Delta \nu ^2}{5 \tilde r^2}-\frac{81 \Delta \nu ^2
   \tilde r^2}{22400}-\frac{\nu ^2 \tilde r^2}{32}+\frac{11 \Delta \nu ^2 \tilde r^4}{179200}+\frac{\nu ^2 \tilde r^4}{768}\; ,
\ee
where here $\vec m=\sigma_0( \nu+\Delta\nu/2,\vec 0_3,\nu-\Delta\nu/2,\vec 0_3)$ and $\tilde r=r/R$.
For $\Delta\nu>0$, negative powers of the separation are present in the exponential, driving the probability to zero at small separations. This behaviour is the same as we observed in BCDP for critical points in 1D density fields.
This suppression on small scales leads to deviations from the $\Delta \nu=0$ case for
\be
r_{1\%}<9600^{1/6}R\Delta\nu^{1/3}\, .
\label{eq:exclscale}
\ee
In the limit of zero separation, the prefactor scales like  $\lim_{r\to 0}A(r)\propto 1/r^{10}$ for $\Delta \nu\neq 0$ and $\lim_{r\to 0} A(r)\propto 1/r^{2}$ for $\Delta \nu= 0$.
In Fig.~\ref{fig:critpoints} we show the numerical and analytical correlation functions of signed critical points as well as the peak correlation function for the same peak height and peak height difference. While the agreement between peaks and critical points is not perfect except on large scales, it is interesting to note that the position of the exclusion scale is in close correspondence.

\section{Numerical Implementation}
\label{sec:MCMC}

\subsection{Fixed Peak Height}
In this section we discuss peaks of fixed (equal and unequal) significance, their small scale exclusion and the large scale bias convergence.

\begin{figure}[t]
\centering
\includegraphics[width=0.49\textwidth]{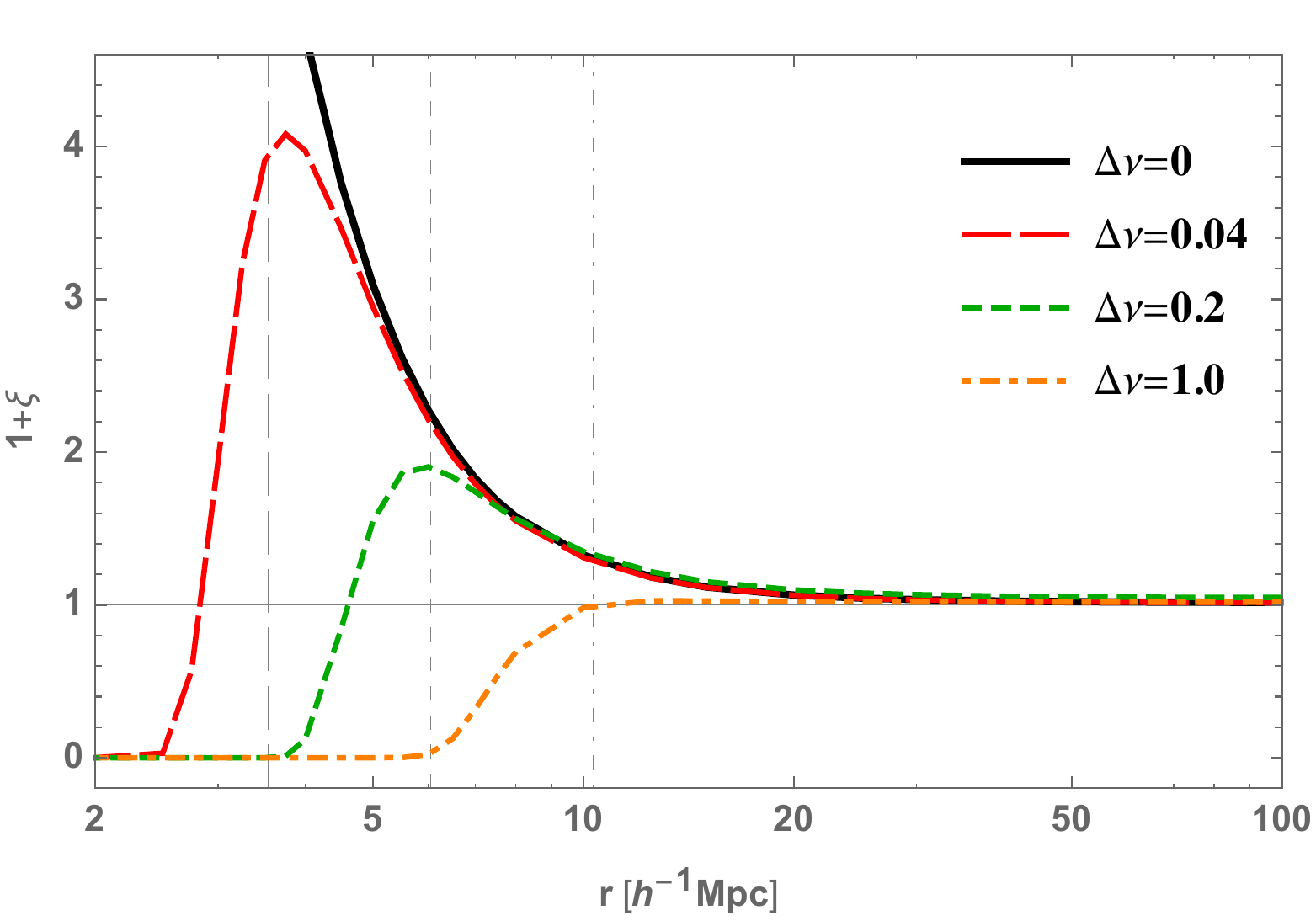}
\caption{
Peak correlation function for a $R=2.2 \hMpc$ smoothed density field with peak heights $\nu_\pm=\bar \nu\pm \Delta \nu/2$. The auto-correlation of $\bar \nu=2$ peaks (black solid) does not show any exclusion. As we increase the separation of peak height, a region of zero probability, i.e. $\xi=-1$ arises at small separations. For larger radii the finite separation correlation function asymptotes to the $\Delta \nu=0$ case. The characteristic scales where this happens can be estimated as $r_{1\%}\approx 9600^{1/6}\Delta \nu^{1/3}R$ and is shown by the vertical lines. \\
}
\label{fig:xivardnu}
\end{figure}

\begin{figure}[t]
\centering
\includegraphics[width=0.49\textwidth]{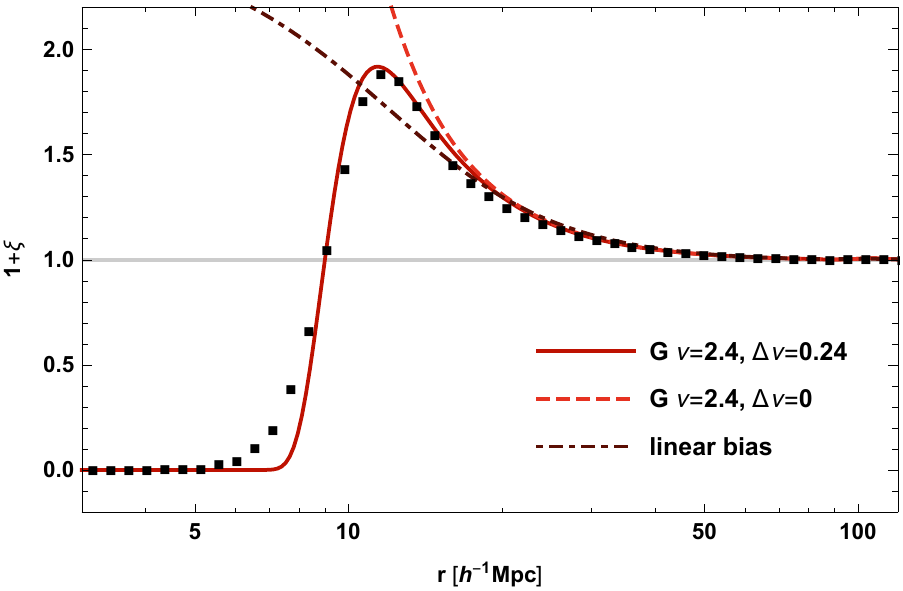}
\includegraphics[width=0.49\textwidth]{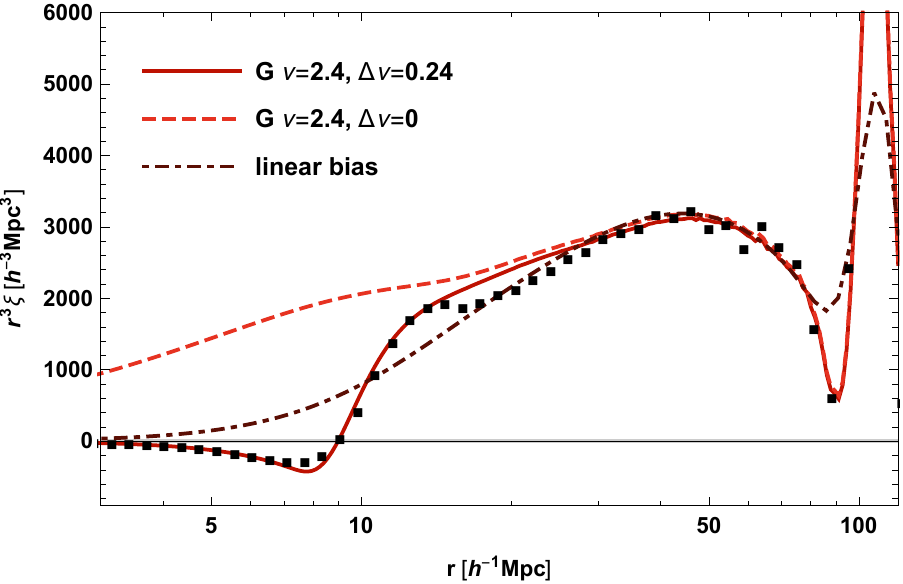}
\includegraphics[width=0.49\textwidth]{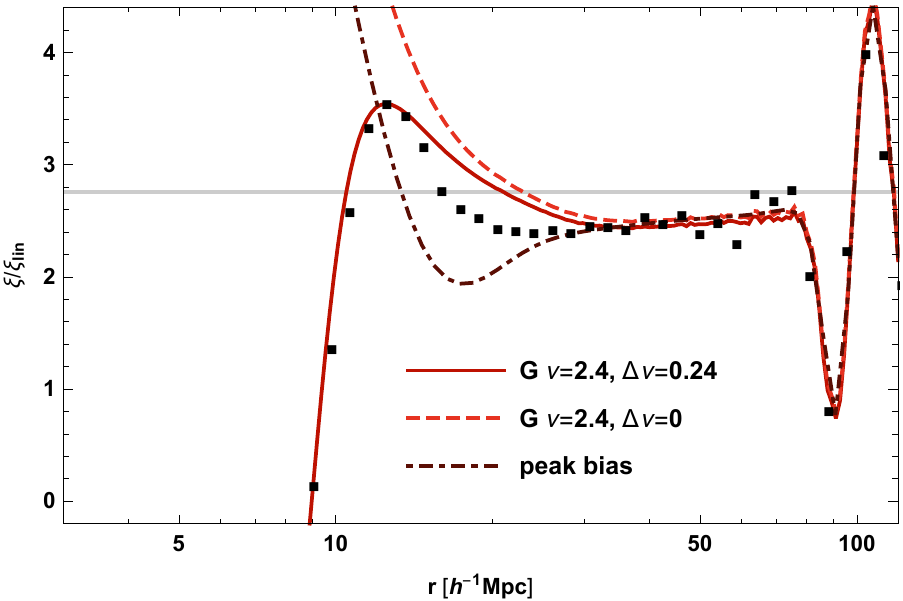}
\caption{Correlation function of bin IV in Lagrangian space modeled by two distinct peak heights with $\bar \nu=2.4,\ \Delta \nu=0.24$ (red solid) and $\bar \nu=2.4,\ \Delta \nu=0$ (red dashed). The points show the measured proto-halo correlation function. \emph{Upper left panel: }Correlation function. We see that the peak correlation function with $\Delta \nu=0$ follows the upturn of the correlation function with respect to linear bias, but does not show exclusion, whereas the $\Delta \nu=0.24$ result does. \emph{Upper right panel: }$r^3$ weighted correlation function, quantifying the contribution to the low-$k$ power spectrum. Regions where the curve is below (above) linear bias lead to negative (positive) stochasticity corrections.  \emph{Lower panel: }Bias with respect to the smoothed linear power spectrum. We clearly see that inside the BAO scale the correlation function is consistently lower than suggested by linear bias (gray horizontal line). This behaviour is correctly reproduced by the linear peak bias in Eq.~\eqref{eq:linpkbias}, which agrees with the full calculation down to separations of $r\approx 30\hMpc$.
}
\label{fig:xibin4}
\end{figure}

The peak correlation function is shown in Fig.~\ref{fig:xivardnu}.
Like peaks in one dimensional density fields, we find that equal height bins ($\Delta \nu=0$) at the same smoothing scale do not exhibit exclusion, and the correlation function keeps growing in the limit $r\to 0$. As we increase the difference in peak significance, the correlation function tends to -1 over an increasing region at small separations. This region corresponds to a vanishing probability of finding two peaks closer than the exclusion radius ($\mathbb P\propto 1+\xi\to 0$). 
On larger scales, the $\Delta \nu\neq 0$ correlation functions asymptote to the $\Delta \nu=0$ case. In this regime, 
the two-point functions entering the covariance matrix are much smaller than the corresponding moments ($\epsilon\sim\xi_{i,l}/\sigma_i^2\ll 1$), so that we can expand the peak correlation function in the small quantity $\epsilon$. 
At leading order, the corresponding peak two-point function is described by the linear scale dependent bias \cite{bbks,regos/szalay:1995,Desjacques:2008jj}
\be
\xi(r;\nu_1,\nu_2)\approx b_{10}(\nu_1)b_{10}(\nu_2)\xi_{0,0}(r)+\left[b_{10}(\nu_1)b_{01}(\nu_2)+b_{01}(\nu_1)b_{10}(\nu_2)\right]\xi_{2,0}(r)+b_{01}(\nu_1)b_{01}(\nu_2)\xi_{4,0}(r)\, ,\label{eq:linpkbias}
\ee
where $\xi_{ij}$ is defined in Eq.~(5) and the bias factors $b_{ij}$ are given by derivatives w.r.t.~the peak height and curvature \cite{Desjacques:2010gz,matsubara:2011,desjacques:2013,matsubara/desjacques:2016},
\be
\sigma_0^i\sigma_2^j b_{ij} = \frac{1}{\bar n} \int\!\!{\rm d}\vec{X}\, \omega(\vec X) \frac{\partial^i}{\partial\nu^i}\frac{\partial^j}{\partial \zeta^j}
\mathbb{P}_\text{1pt}(\vec X)\, .
\ee
This linear bias model differs from the usual scale independent linear bias model ($\xi=b_1^2 \xi_\text{lin}$) due to the fact that $\xi_{0,0}$ contains an explicit smoothing scale, and the presence of the higher derivative terms ($\xi_{2,0},\xi_{4,0}$) which enhance the BAO \cite{Desjacques:2008jj} (see the bottom panel of Fig.~\ref{fig:xibin4}). We can also calculate the next and next-to-next-to-leading order corrections as done in \cite{Desjacques:2010gz,Matsubara:2019sab} and, more systematically, from the peak perturbative bias expansion \cite{desjacques:2013,Lazeyras:2015giz}. However, let us stress that these bias expansions converge very slowly and, for any realistic perturbative order, cannot capture the peak of the correlation function just outside the exclusion zone and even less so the very non-linear exclusion itself (as emphasized by the one dimensional analysis of \cite{Baldauf:2015fbu}). In this regime, at low and intermediate separations, our full numerical integration is mandatory.


\begin{table}[t]
\begin{tabular}{cccccc}
\hline
\hline
Bin&$R\ [\hMpc]$&$\bar \nu$&$\Delta \nu$ & $b_1$ & $M\ [h^{-1}M_\odot]$\\
\hline
I & 1.68 &1.34 &0.36&0.11&$7.73 \times 10^{12}$\\
II & 2.2 &1.7 &0.36&0.35&$2.33 \times 10^{13}$\\
III & 3.1 &2 &0.36&0.82&$6.92 \times 10^{13}$\\
IV & 4.3 &2.4 &0.36&1.65&$2.01 \times 10^{14}$\\
V & 6.3 &2.9 &0.36&3.17&$5.68 \times 10^{14}$\\
\hline
\hline
\end{tabular}
\caption{Properties of the five proto-halo mass bins constructed from FoF haloes found at $z=0$, namely Gaussian smoothing scale, mean peak height, peak height root mean square scatter, Lagrangian bias and mean mass.}
\label{tab:massbins}
\end{table}

\begin{figure}[t]
\includegraphics[width=0.49\textwidth]{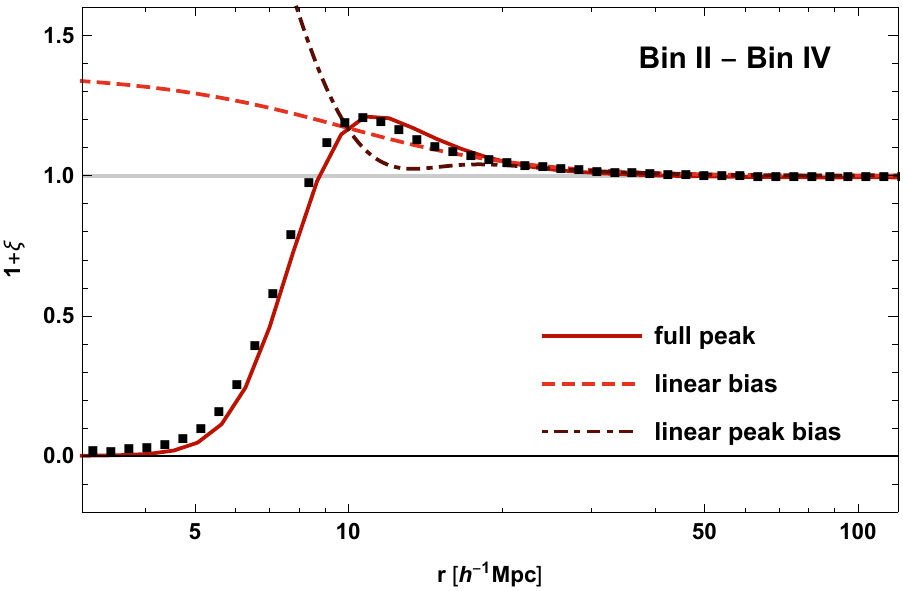}
\includegraphics[width=0.49\textwidth]{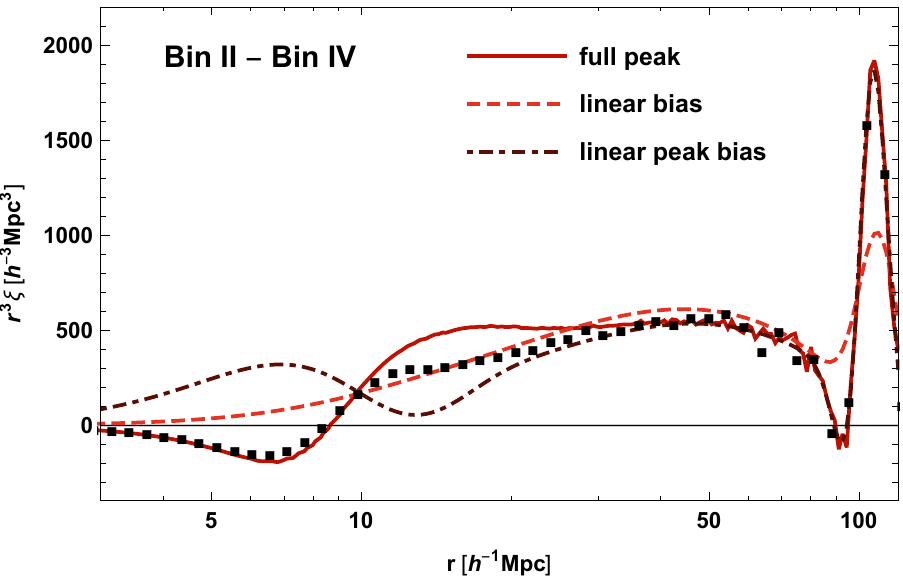}
\caption{Cross correlation function between mass bins II and IV. \emph{Left panel: }In the correlation function we see that the choice of the smoothing radii and peak heights correctly captures the exclusion scale observed for the proto-halo cross-correlation in the simulations. \emph{Right panel: }Correlation function weighted by $r^3$, highlighting the contributions to the low-$k$ power spectrum amplitude. The regions that are below (above) the linear bias curve lead to negative (positive) shot noise corrections.}
\label{fig:xibin24}
\end{figure}

\subsection{Peak Height Bins}
Eventually we want to use peaks as a proxy for haloes. In order to estimate the agreement between the peak model predictions and the properties of actual haloes, we consider a suite of 16 Gadget $N$-body simulations initialized at redshift $z=99$ using second order Lagrangian perturbation theory (see \cite{Baldauf:2014fza} for more details). Haloes are identified using a Friends-of-Friends (FoF) halo finder in the $z=0$ non-linear density field. Their constituent particles can be traced back to the simulation initial conditions - or Lagrangian space - to identify the halo progenitor - or proto-halo - whose centre of mass gives the proto-halo (Lagrangian) position.
We bin the whole halo distribution into five mass bins, each of which is spanning a factor of three in mass. The properties of these halo catalogues are summarized in Tab.~\ref{tab:massbins}.

As we have seen above, the peak model predictions depend on the peak height $\nu$ and the smoothing scale $R$. In this study we consider haloes (and consequently proto-haloes) in a number of mass bins and each mass in the bin would correspond to a different filter scale. We have found in \cite{Baldauf:2014fza} that the cross-correlation between proto-haloes and the Gaussian density field can be reasonably well described by a single Gaussian filtering scale. When quantifying the properties of the underlying density field at the proto-halo position, we will thus filter the initial density field with a Gaussian filter at the scale fitted in \cite{Baldauf:2014fza} (which was based on the same simulations and halo sample) and quoted in Tab.~\ref{tab:massbins}. We have ascertained that a change in the width of the mass bins (at fixed central mass) does not change the extent of the exclusion scale in the proto-halo correlation function. This is consistent with the observation that the scatter of the threshold does not significantly depend on the width of the mass bin \cite{Ludlow:2011tpf,paranjape/etal:2013}.

For such a fixed filter scale we find that the observed distribution of peak heights, i.e.~linear overdensities at the peak location, is approximately Gaussian, as we show in App.~\ref{app:scatter}. 
Hence, our findings for the distribution of actual proto-halo peak heights would suggest to sample the peak height from a Gaussian. To simplify the modelling and accelerate the evaluation of the peak correlation function, we chose instead to implement a finite peak height difference even for the auto-correlation of peaks in the same mass bin.\footnote{If the distribution of peak heights in the bins is Gaussian $\nu\sim \mathcal{N}(\bar \nu,\sigma)$ then the difference between the peak heights is a Gaussian  $\nu_2-\nu_1\sim \mathcal{N}(\Delta \nu=\bar \nu_2-\bar \nu_1,\sigma_\Delta=\sqrt{\sigma_1^2+\sigma_2^2})$. The mean of the absolute value of the peak height difference is then given by $2/\sqrt{\pi} \sigma\approx 1.12 \sigma$ for a single bin and 
\be
\overline{\Delta \nu}= \sqrt{\frac{2}{\pi }} \sigma _{\Delta } \exp\left({-\frac{\Delta \nu ^2}{2 \sigma _{\Delta }^2}}\right)+\Delta \nu\  \text{erf}\left(\frac{\Delta \nu }{\sqrt{2}
   \sigma _{\Delta }}\right)
\ee
for distinct bins. The single-bin result is recovered for $\sigma_\Delta \gg \Delta \nu$.} We show the results of this implementation of the peak model in comparison to measurements of the proto-halo correlation function in Fig.~\ref{fig:xibin4}. We see that the $\Delta \nu \neq 0$ case reproduces the exclusion scale but overpredicts the correlation function just outside the exclusion scale at $R\approx 20\hMpc$. The bottom panel of Fig.~\ref{fig:xibin4} shows that, for $r> 40\hMpc$, the full peak correlation functions with and without exclusion match each other and also agree with the linear peak correlation function. Notice that the proto-halo correlation function has an amplitude lower than $b_{10}^2\xi_\text{lin}$ (linear bias) indicated by the horizontal gray line. In particular, the linear, scale-independent biasing $b_{10}^2\xi_\text{lin}$ is only approached beyond the BAO scale and not within it. Assuming linear bias within the BAO scale might thus lead to biased estimates of the amplitude of fluctuations. The mismatch between the peak and proto-halo correlation functions just outside of the exclusion scales might be related to the peak selection function being more complicated than the Gaussian filter employed here. We have explored sampling from the actual Gaussian peak height distribution and will discuss the results in Sec.~\ref{sec:realizations} below. While the Gaussian sample of peak heights does show exclusion, the transition between the continuous and excluded regions is significantly smoother than what is observed for proto-haloes and the $\Delta \nu\neq 0$ peak sample.

We show the cross-correlation between proto-halo mass bins II and IV in Fig.~\ref{fig:xibin24}. In this cross-correlation setting, the individual smoothing scales differ and so do the peak heights. 
A pronounced exclusion region is also found in that case both in the measurements and in our modeling based on peak correlation functions with different height and smoothings. The {predicted} size of the exclusion region agrees with the measurements of the proto-halo cross-correlation function, but there is up to 20\% discrepancy just outside the exclusion zone, {after the maximum of the correlation function at roughly $10 \hMpc$. Here, the measured correlations are found to lie above the linear bias prescription but below the peak model}. These are probably due again to our filter not properly describing the proto-halo selection function as well as tidal effects or the impact of the FoF halo finder. 

\subsection{Shotnoise Corrections}
Small scale exclusion is relevant for the large-scale (low-$k$) corrections to the halo stochasticity in the power spectrum \cite{smith/etal:2007,Baldauf:2013hka}. To see how this sensitivity arises, let us consider the expression of the halo power spectrum in terms of the correlation function
\be
P_\text{hh}(k)=\frac{1}{\bar n}+4\pi \int \derd \ln r\; r^3 \xi_\text{hh}(r) j_0(k r)\; .
\ee
For small wavenumbers the above expression simplifies to
\be
P_\text{hh}(k\to0)\approx \frac{1}{\bar n}+4\pi \int \derd \ln r\; r^3 \xi_\text{hh}(r)\;.
\ee
Thus, the low-$k$ power spectrum is just the $r^3$-weighted logarithmic integral over the correlation function augmented by the Poisson noise $1/\bar n$. The integral arises from the contribution of distinct pairs, while the Poisson noise corresponds to ``self-pairs''.
For the linear correlation function (and linearly biased versions of it) the above integral vanishes. This changes once higher order perturbative corrections or exclusion corrections are taken into consideration.

Instrumental for the understanding of stochasticity is the ability to describe the integrand in the above equation, i.e., $r^3 \xi$. We show this $r^3$-weighted correlation function in the top right panel of Fig.~\ref{fig:xibin4}, which emphasizes that both the peak and proto-halo correlation functions considerably deviate from linear biasing at small separations.
The regions where the full peak or halo correlation function lies below the linear bias curve lead to a negative stochasticity correction on large scales,
while the part of the curve that lies above leads to a positive stochasticity correction. Depending on which of the two effects dominates, the overall stochasticity correction can be either positive or negative. Generally, we can remark that the peak model captures the behaviour of the $r^3$-weighted correlation function quite well.

\subsection{Velocities}
\begin{figure}[t]
\includegraphics[width=0.49\textwidth]{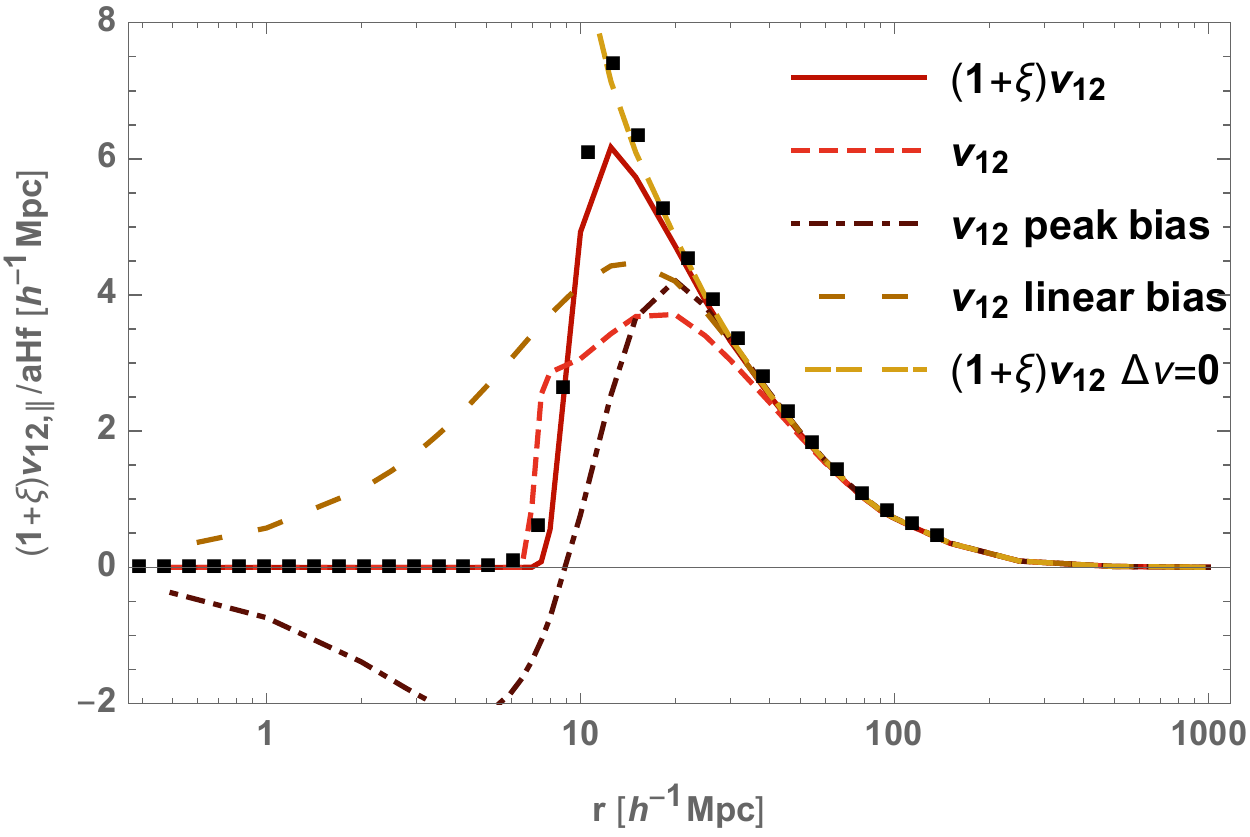}
\caption{
Mass weighted velocity correlation (mean infall) of peaks with smoothing scale $R=4.3\hMpc$ compared to the measured velocity correlator for proto-haloes in bin IV of the simulations. The agreement is fairly good except for a slightly too low amplitude of the peak predictions at the peak of the mean infall.
Clearly the velocity correlator inherits the exclusion from the density correlation function. The linear bias works down to separations of $30\hMpc$. }
\label{fig:vel}
\end{figure}

\begin{figure}[t]
\includegraphics[width=0.49\textwidth]{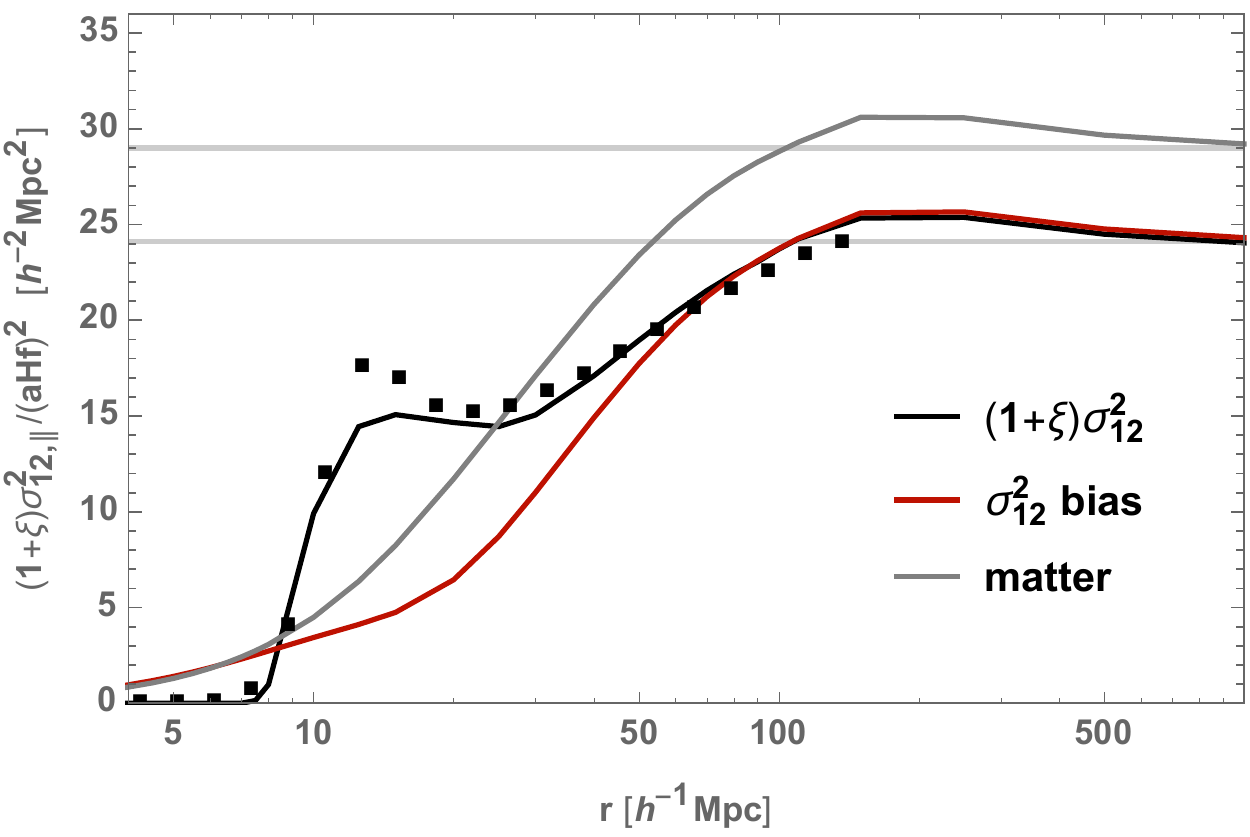}
\includegraphics[width=0.49\textwidth]{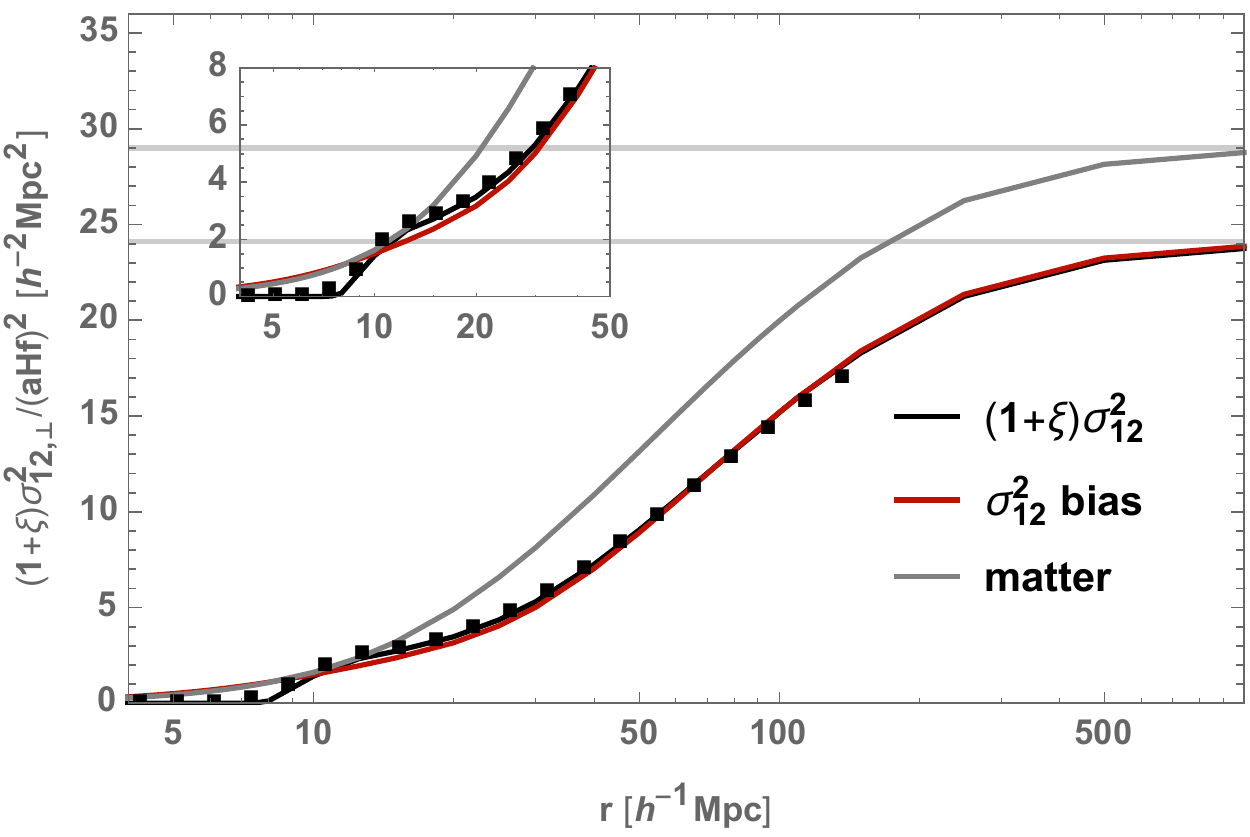}
\caption{Mass weighted velocity dispersion parallel (\emph{left panel}) and perpendicular (\emph{right panel}) to the separation vector for mass bin IV. We show the simulation measurement as black squares and the predictions of the peak model (black), linear velocity bias (red) and the underlying matter distribution (gray). Due to the velocity bias and smoothing, there is an offset between the large-scale limit of the linear matter and halo velocity dispersion, i.e., $\sigma_v^2$. On small scales there are even more significant deviations from the linear matter dispersion.}
\label{fig:veldisp}
\end{figure}

Having studied the clustering of peaks and their correspondence to proto-halo positions in Lagrangian space, let us now focus on their velocities. For the proto-haloes we define the velocity as the mean velocity of their constituent dark matter particles. In the Zeldovich approximation \cite{zeldovich:1970}, the Lagrangian velocity is directly related to the halo displacement ${\vec \Psi=\vec v/\mathcal{H}f}$ and thus determines the position of the haloes in Eulerian space. We will make use of this fact below in Sec.~\ref{sec:evol}. The statistics of the Zeldovich displacement is straightforward to implement since it is the anti-derivative of the linear density field due to Poisson equation $\vec \Psi(k)=-\imath \vec k/k^2 \delta_\text{s}(k)$. As such, the joint statistics of $\vec\Phi=(\vec X,\vec \Psi)$ follows a Gaussian with zero mean and covariance matrix
\be
\mathbf{C}=
\begin{pmatrix}
\la \vec X\!\cdot\! \vec X^{\rm T}\ra & \la \boldsymbol X \!\cdot\! \boldsymbol\Psi^{\rm T}\ra\\
\la \boldsymbol\Psi\!\cdot\!\boldsymbol X^{\rm T}\ra & \la \boldsymbol\Psi\!\cdot\!\boldsymbol\Psi^{\rm T }\ra
\end{pmatrix}
=
\begin{pmatrix}
\mathbf{C}_{ {\vec X,\vec X}} & \mathbf{C}_{{\vec X,\vec \Psi}}\\
\mathbf{C}_{{\vec X,\vec \Psi}}^{\rm T}&\mathbf{C}_{ {\vec \Psi,\vec \Psi}}
\end{pmatrix}
=
\begin{pmatrix}
\mathbf{\Omega}_{ {\vec X,\vec X}} & \mathbf{\Omega}_{{\vec X,\vec \Psi}}\\
\mathbf{\Omega}_{{\vec X,\vec \Psi}}^{\rm T}&\mathbf{\Omega}_{{\vec \Psi,\vec \Psi}}
\end{pmatrix}^{-1}
=\mathbf\Omega^{-1}\,,
\ee
where $\mathbf\Omega=\mathbf C^{-1}$ is the precision matrix.

In contrast to the 1D case considered in BCDP, the three dimensional velocities have two components, with different statistical properties: one along the separation of the peaks and one transverse to it. Let us start by considering the mean streaming velocity along the separation axis $\vec {\hat r}_{12}$
\be
v_{12,\parallel}=\frac{\left\langle (\vec v_2-\vec v_1)\scalp \hat{\vec r}_{12}(1+\delta_\text{pk,1})(1+\delta_\text{pk,2})\right\rangle}{1+\xi}=\frac{\la (\vec v_2-\vec v_1)\scalp \hat {\vec r}_{12}w(\boldsymbol X)\ra }{\la w(\boldsymbol X) \ra},
\ee
where $w(\boldsymbol X)$ is the peak condition at both locations and $\la w(\boldsymbol X) \ra=(1+\xi_\text{pk})\bar n_\text{pk}^2$.
For convenience, and to make connections to the displaced peaks discussed below, we express velocities in units of displacements. Performing the Gaussian integral over the velocity components of the state vector, we thus have
\be
\label{eq:v12}
\begin{split}
(1+\xi_\text{pk})\frac{v_{12{,\parallel}}}{\mathcal{H}f}=&\frac{\left\langle (\vec v_2-\vec v_1)\scalp \hat{\vec r}_{12}(1+\delta_\text{pk,1})(1+\delta_\text{pk,2})\right \rangle}{\mathcal H f} \\
=& -\frac{1}{\bar n_\text{pk}^2}\!\int\! \derd \boldsymbol X\; \boldsymbol X^{\rm T} \!\!\cdot\!  \mathbf{C}_{ {X}} ^{-1}  \!\!\cdot\! \mathbf{C}_{{\!X\!\Psi}} \!\!\cdot\! \,\boldsymbol U_{{\parallel}}^{{\rm T}}\; \frac{w(\boldsymbol X)}{\sqrt{(2\pi)^6 \det \mathbf{C}_{ {X}} }} \eh{-\frac12 \boldsymbol X^{\rm T}\!\!\cdot\! \mathbf{C}_{ {X}} ^{-1} \!\!\cdot\!\boldsymbol X}\; .
\end{split}
\ee
Here, the velocity difference is expressed in terms of the state vector of the displacement at the two positions, $\vec \Psi=(\vec \Psi_1,\vec \Psi_2)$ using a linear transformation
\be
\label{eq:U}
\vec\Psi_2-\vec \Psi_1
=
\vec U\scalp \vec \Psi\; ,
\ee
where
\be
\vec U=
\begin{pmatrix}1 & 0& 0&-1 & 0& 0\\0&1 & 0& 0&-1 & 0\\0&0&1 & 0& 0&-1 \end{pmatrix}\; ,
\ee
and $\vec U_{\parallel}=\vec {\hat r}_{12}\scalp \vec U$.
The Gaussian integral was performed using the relation
$\det \mathbf{C}=\det\mathbf{C}_{ {X}}/\det \mathbf{\Omega}_{ \Psi} $ together with
\be
\vec\Phi^{\rm T}\scalp
\mathbf{ C}^{-1}\scalp\vec\Phi=
\boldsymbol X^{\rm T}\cdot \mathbf{C}_ {X} ^{-1} \cdot\boldsymbol X
+(\vec \Psi-\vec\mu)^{\rm T}\cdot\vec \Omega_{\Psi} \cdot (\vec \Psi-\vec\mu),
\ee
where 
$\vec \mu^{\rm T}=-\vec X^{\rm T}\scalp\mathbf{\Omega}_{{X\Psi}}\scalp\mathbf{\Omega}_{\Psi}^{-1}=\vec X^{\rm T}\scalp\mathbf{C}_{X}^{-1}\scalp\mathbf{C}_{{X\Psi}}$.

In the large separation limit, where $\epsilon\sim\xi_{i,l}/\sigma_i^2\ll 1$, we can expand to linear order in $\epsilon$ and recover the linear velocity bias in the peak model \citep{Desjacques:2008jj,desjacques/sheth:2010}
\be
\label{eq:v12bias}
\frac{v_{12{,\parallel}}}{\mathcal{H}f}\approx b_{10}\left(\xi_{-1,1}-\frac{\sigma_0^2}{\sigma_1^2}\xi_{1,1}\right)+b_{01}\left(\xi_{1,1}-\frac{\sigma_0^2}{\sigma_1^2}\xi_{3,1}\right),
\ee
where the bias coefficients are defined in Appendix~\ref{sec:bias}.
Note that this has a richer structure than the velocity bias in the local bias model, which would only yield the term proportional to $\xi_{-1,1}$ (and in most cases this term even lacks the explicit smoothing used here).

The mean relative displacement is shown in Fig.~\ref{fig:vel}, the red solid line being the full numerical implementation of Eq.~(\ref{eq:v12}) which is the novelty of this work. For comparison, we also display the linear and first order peak prediction given by Eq.~(\ref{eq:v12bias}). Like the density correlator, there is a pronounced small scale exclusion both in the model and the data. The relative infall has to go to zero on small scales as there are no pair closer than the exclusion scale. The linear bias predictions (both with and without peak corrections) fail at roughly $30\, \hMpc$. Below this scale only the full peak calculation is in close agreement with the data, capturing both the maximum of the mean relative velocity at roughly $10-20 \hMpc$ and the exclusion zone with zero mean mass weighted relative velocity at low separation.

The velocity dispersion along the separation and perpendicular to the separation are defined respectively as
\begin{align}
\label{eq:sig12pa}
\sigma_{12,\parallel}^2=&\frac{\la \left[(\vec v_2-\vec v_1)\scalp \hat {\vec r}_{12}\right]^2(1+\delta_\text{pk,1})(1+\delta_\text{pk,2})\ra}{1+\xi_\text{pk}}\,,\\
\sigma_{12,\perp}^2=&\frac{\la \left[(\vec v_2-\vec v_1)-(\vec v_2-\vec v_1)\scalp \hat {\vec r}_{12}\, \hat {\vec r}_{12}\right]^2(1+\delta_\text{pk,1})(1+\delta_\text{pk,2})\ra}{1+\xi_\text{pk}},
\end{align}
such that for instance (and equivalently for the perpendicular component if $U_{\parallel}$ is replaced by $U_{\perp}$)
\be
\label{eq:sig12pe}
(1+\xi_\text{pk})\frac{\sigma_{12,\parallel}^{2}}{(\mathcal{H}f)^2}=\frac{1}{\bar n_\text{pk}^2}\!\int \!\derd \boldsymbol X\; \Bigl[\left(\boldsymbol X^{\rm T} \!\!\cdot\!  \mathbf{C}_{ {X}} ^{-1}  \!\!\cdot\! \mathbf{C}_{{\!X\!\Psi}} \!\!\cdot\! \,\boldsymbol U_{{\parallel}}^{{\rm T}}\right)^2+{U}_{{\parallel}}\!\!\cdot\!  \mathbf{\Omega}_{\boldsymbol {\Psi}}^{-1} \!\!\cdot\! \, \boldsymbol U_{{\parallel}}^{{\rm T}}\Bigr]\; \frac{w(\boldsymbol X)}{\sqrt{(2\pi)^6 \det \mathbf{C}_{\boldsymbol {X}} }} \eh{-\frac12 \boldsymbol X^{\rm T}  \!\!\cdot\!  \mathbf{C}_{\boldsymbol {X}} ^{-1}  \!\!\cdot\! \boldsymbol X} \!.
\ee
Eventually, at leading order, we get \cite{desjacques/sheth:2010}
\be
\label{eq:sig12pebias}
\frac{\sigma_{12,\perp}^2}{(\mathcal{H}f)^2}\approx \frac{2}{3} \sigma_{v,\text{pk}}^2 - 
 \frac{2}{3} \left[\xi_{-2,0} + \xi_{-2,2} - 
    2 \frac{\sigma_0^2}{\sigma_1^2} (\xi_{0,0} + \xi_{0,2}) + \frac{\sigma_0^4}{\sigma_1^4} (\xi_{2,0} + \xi_{2,2})\right],
\ee
\be
\label{eq:sig12pabias}
\frac{\sigma_{12,\parallel}^2}{(\mathcal{H}f)^2}\approx  \frac{1}{3} \sigma_{v,\text{pk}}^2 - 
 \frac{1}{3} \left[\xi_{-2,0} -2 \xi_{-2,2} - 
    2 \frac{\sigma_0^2}{\sigma_1^2} (\xi_{0,0} -2 \xi_{0,2}) + \frac{\sigma_0^4}{\sigma_1^4} (\xi_{2,0} -2 \xi_{2,2})\right],
\ee
where $\sigma_{v,\text{pk}}^2\equiv \sigma_{-1}^2-\sigma_0^4/\sigma_1^2$.

The peak and halo displacement dispersion are shown in Fig.~\ref{fig:veldisp}. Again, we display both the full numerical calculation in black as given by Eq.~(\ref{eq:sig12pa}-\ref{eq:sig12pe}), the novelty of this work, and the first order peak prediction of Eq.~(\ref{eq:sig12pebias}-\ref{eq:sig12pabias}) in red. On large scales the displacement dispersions of matter (gray) and peaks (black) deviate due to the explicit smoothing scale in the peak displacement dispersion and due to explicit velocity bias effects. The measured halo displacement dispersions follow the prediction of the peak model down to small separations. The linear peak bias prediction provides a good description of the full peak dispersion down to separations of $40\,\hMpc$ (red) but fails to predict the bump in the velocity dispersion between $10$ and $20 \hMpc$ -- notably parallel to the separation -- together with the exclusion at small separations which are both well captured by the full peak calculation. The exact amplitude of the bump shows some difference between the haloes and the peak model, similarly to the density and relative velocity correlators. 

\section{Evolution to Eulerian Space}
\label{sec:evol}
The strategy of evolving the 3D peaks to Eulerian space closely follows the steps laid out in BCDP, but we will spell out the important steps for the readers convenience. In particular, we will consider the Zeldovich displacement of a peak according to the initial velocity field at the peak location. At the perturbative level this calculation was performed in \cite{Desjacques:2010gz,Baldauf:2016aaw}.
The motivation for using the Zeldovich approximation for displacing the haloes is two-fold. First, haloes are extended objects and in the model we are working with they are patches of conserved mass whose center of mass is simply moving from their Lagrangian to their Eulerian position while the mass distribution collapses around this center of mass. This kind of objects is particularly amenable to a perturbative treatment since they never experience shell-crossing. The second reason is computational convenience. In the Zeldovich approximation the displacement field is linear in the underlying field and thus Gaussian (the inferred density is not). This allows us to work with the Gaussian multipoint-PDF of the field, field derivatives and displacements.
\begin{figure}[t]
\centering
\includegraphics[width=0.49\textwidth]{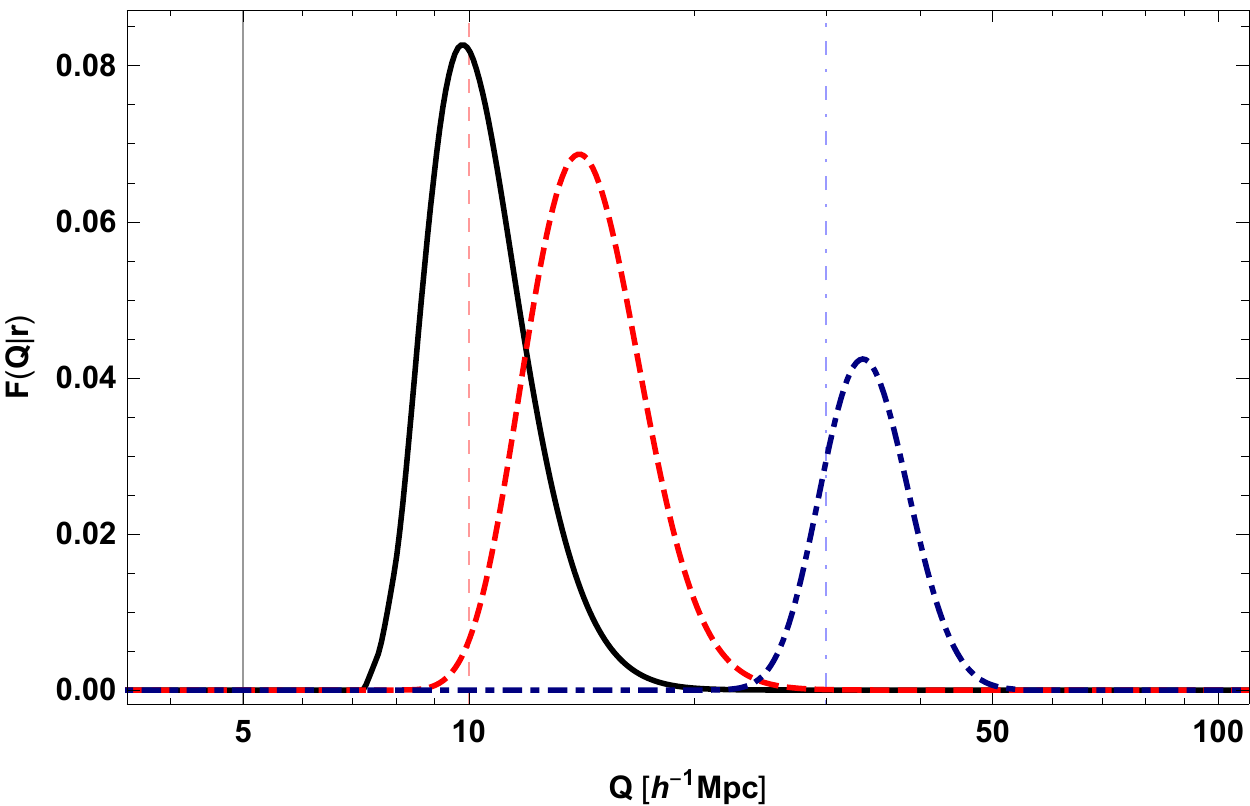}
\caption{Integrand $F(Q|r)$ of Eq.~\eqref{eq:lagconv}. The vertical lines indicate the Eulerian scale $r$, and the offset between this scale and the center of the Gaussian of the same color is given by the mean infall $\la v_{12}\ra/\mathcal{H}f$. }
\label{fig:disppeakcontrib}
\end{figure}

In this description, the number density of Eulerian peaks reads
\newcommand{\ddir}{\delta^\text{(D)}}
\be
\begin{split}
1+\delta_\text{pk}(\vec r)=&\frac{1}{\bar n_\text{pk}}\sum_\text{pk} \ddir\left(\vec r- \vec r_\text{pk}\right)=\int \derd^3 q' \ddir\left[ \vec r- \vec q'- D_{\!+}\vec \Psi(\vec q')\right]\sum_\text{pk}\ddir( \vec q'- \vec q_\text{pk})\,,\\
=&\int \derd^3 q' \int \frac{\derd^3 k}{(2\pi)^3}\eh{\ii  \vec k\scalp( \vec r-\vec  q')}w(\vec X)\eh{-\ii  D_{\!+}  \vec k \scalp\vec \Psi( \vec q')}\,,
\end{split}
\ee
and their correlation function 
\be
\xi(r)=\la\delta_\text{pk}(\vec 0)\delta_\text{pk}(\vec r)\ra=\frac{1}{\bar n_\text{pk}^2}\int \derd^3 Q  \int \frac{\derd^3 k}{(2\pi)^3}  \eh{\ii \vec k (\vec Q-\vec r)}\Bigl\langle \eh{-\ii  D_{\!+} \vec k\scalp(\vec \Psi_1-\vec \Psi_2)}  w(\vec X_1)w(\vec X_2)\Bigr\rangle-1\; , \label{eq:xivel}
\ee
where $\vec Q=\vec q_2-\vec q_1$ is the Lagrangian separation of the peaks, $\vec \Psi_1$ and $\vec \Psi_2$ are the halo displacements at the respective positions and $D_+$ is the amplitude of the growing mode in the linear regime of structure formation.
Using Eq.~(\ref{eq:U}), then Eq.~\eqref{eq:xivel} yields
\begin{align}
\xi(r)=&\frac{1}{\bar n_\text{pk}^2}\int \derd^3 Q  \int \frac{\derd^3 k}{(2\pi)^3} \int \derd \boldsymbol X  \eh{\ii \vec k \scalp (\vec Q-\vec r)}\nonumber\\
&\times\frac{w(\boldsymbol X)}{\sqrt{(2\pi)^6 \det \mathbf{C}_{ {X}} }}
\eh{-\frac12 \boldsymbol X^{\rm T}\!\!\cdot\! \mathbf{C}_{ {X}} ^{-1} \scalp \boldsymbol X-\frac12 D_{\!+}^2 k^2 \,\boldsymbol U \scalp \mathbf{\Omega}_{ {\Psi}}^{-1}\scalp \vec U^{{\rm T}}-\ii D_{\!+} k \vec X^{\rm T} \scalp \mathbf{C}_{ {X}} ^{-1} \scalp \mathbf{C}_{{\!X\!\Psi}} \scalp \vec U^{{\rm T}}
}
-1\, .
\label{eq:movedpeak1d}
\end{align}
The Gaussian integral over wavenumbers, $k$, can be trivially performed and leaves us with a convolution
\begin{equation}
\la\delta_\text{pk}(\vec 0)\delta_\text{pk}(\vec r)\ra=\int \derd^3 Q\ F(\vec Q|\vec r)-1\,,
\label{eq:lagconv}
\end{equation}
where
\be
\begin{split}
F(\vec Q|\vec r)=\frac{1}{\bar n_\text{pk}^2} \! \int\! \derd \boldsymbol X &\frac{w(\boldsymbol X)}{\sqrt{(2\pi)^{20} \det \mathbf{C}_{ {X}} }}\eh{-\frac12 \boldsymbol X^{\rm T}  \scalp \mathbf{C}_{ {X}} ^{-1}  \scalp \boldsymbol X}\\
\times & \frac{1}{\sqrt{(2\pi)^3 \det \vec \Sigma}}\eh{-\frac12
(\vec Q-\vec r-\vec{\bar \Psi})^{\rm T}\scalp\vec\Sigma^{-1}\scalp(\vec Q-\vec r-\vec{\bar  \Psi})},
\end{split}
\ee
\be\label{eq:peakconv}
\mathrm{with}\quad\vec{\bar\Psi}=D_{\!+} \boldsymbol X^{\rm T} \!\!\cdot\!  \mathbf{C}_{ {X}} ^{-1}  \!\!\cdot\! \mathbf{C}_{{\!X\!\Psi}} \!\!\cdot\! \,\boldsymbol U^{{\rm T}}\,,\quad \mathrm{and}\quad
\vec \Sigma=D_{\!+}^2 \boldsymbol U\!\!\cdot\!\mathbf{\Omega}_{ {\Psi}}^{-1}\!\!\cdot\!\boldsymbol U^{{\rm T}}\,.%
\ee
This expression can be compared to the corresponding expression for the Zeldovich correlation function for dark matter,
for which the mean displacement vanishes $\bar \Psi=0$.

\begin{figure}[t]
\centering
\includegraphics[width=0.49\textwidth]{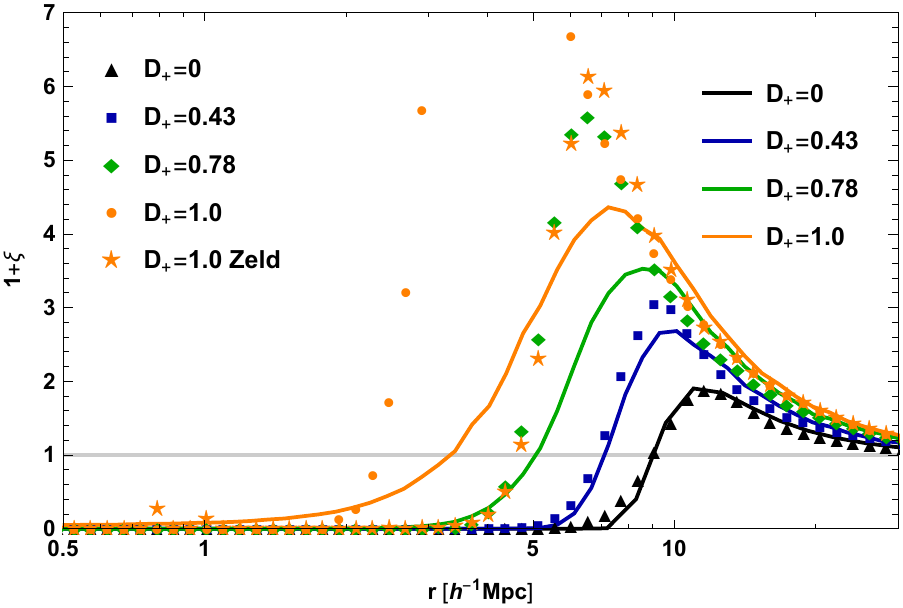}
\includegraphics[width=0.49\textwidth]{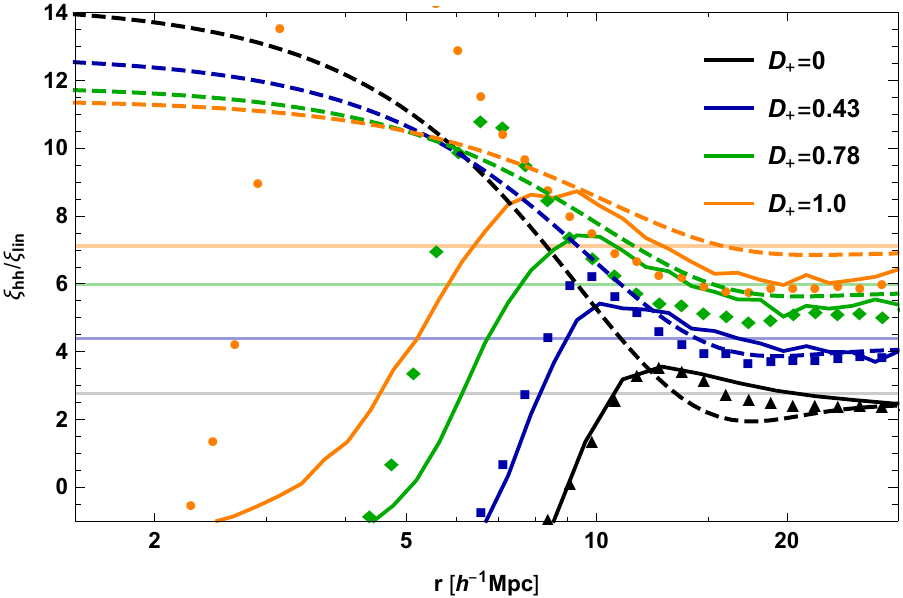}
\caption{Correlation function of Zeldovich evolved peaks (\emph{left panel}) and ratio of peak/halo and matter correlation function (\emph{right panel}). The lines show the peak correlation function corresponding to various redshifts and the points show the measured halo correlation function in the initial conditions and today. The evolved peaks capture the reduction of the extent of the exclusion zone.
We show the correlation function of haloes identified at $z=0$ traced back to $z=0.5$, $z=1$ and $z_\text{i}=99$. The stars in the left panel show the correlation function of the $z_\text{i}=99$ proto-haloes displaced by the actual mean Zeldovich displacement field of its constituent particles. 
In the right-hand panel, we show the ratio of the evolved peak correlation function to the linear dark matter correlation and superimpose the linear prediction from Eq.~(\ref{eq:evolvedpkbias}) (dashed lines) in addition to the full calculation given by Eq.~(\ref{eq:lagconv-2}) (solid lines). Note that both the prediction and the proto-halo clustering is below the horizontal linear bias prediction.
}
\label{fig:disppeak}
\end{figure}
Defining $\mu=\vec{\hat r}\cdot\vec{\hat Q}$ as well as $\Sigma_{ij}=\Sigma_\perp(\delta_{ij}^\text{(K)}-\hat Q_i \hat Q_j)+\Sigma_\parallel\hat Q_i\hat Q_j$ and 
 $\Sigma_{ij}^{-1}=\Sigma_\perp^{-1}(\delta_{ij}^\text{(K)}-\hat Q_i \hat Q_j)+\Sigma_\parallel^{-1}\hat Q_i\hat Q_j$, we get
\be
\begin{split}
 (Q-r-\bar \Psi)_i\Sigma^{-1}_{ij}(Q-r-\bar \Psi)_j =&(Q-\bar \Psi)^2 \Sigma_\parallel^{-1}+r^2 \Sigma_\perp^{-1}\\
 +&2(\bar \Psi \Sigma_\parallel^{-1}-Q \Sigma_\parallel^{-1})r\mu+(\Sigma_\parallel^{-1}-\Sigma_\perp^{-1})r^2\mu^2.
\end{split}
\ee
The angular integral can be performed analytically using
\be
\int_{-1}^1 \derd \mu \eh{-\frac12 \alpha \mu^2+\beta \mu+\gamma}=
\eh{\frac{\beta^2}{2 \alpha} + 
  \gamma} \sqrt{\frac{\pi}{2\alpha}} \left(\text{erf}\left[\frac{\alpha - \beta}{\sqrt{2 \alpha}}\right] + 
   \text{erf}\left[\frac{\alpha + \beta}{\sqrt{2 \alpha}}\right]\right)\, ,
\ee
leaving only one numerical integral in equation~\eqref{eq:lagconv} over the magnitude of the Lagrangian separation $Q$. Upon performing the angular integration, we arrive at
\begin{equation}
\la\delta_\text{pk}(\vec 0)\delta_\text{pk}(\vec r)\ra=4\pi\int_0^\infty \derd Q\,Q^2\, F( Q| r)-1\,,
\label{eq:lagconv-2}
\end{equation}
where now
\begin{gather}
\label{eq:F}
F( Q|r)= \frac{1}{\sqrt{(2\pi)^3 \det \vec \Sigma}}
\eh{\frac{\beta^2}{2 \alpha} + 
  \gamma} \sqrt{\frac{\pi}{2\alpha}} \left(\text{erf}\left[\frac{\alpha - \beta}{\sqrt{2 \alpha}}\right] + 
   \text{erf}\left[\frac{\alpha + \beta}{\sqrt{2 \alpha}}\right]\right) \\
   \mbox{with}\qquad \alpha = \big(\Sigma_\parallel^{-1}-\Sigma_\perp^{-1}\big) r^2 \;,
   \quad \beta = \big(Q-\bar\Psi\big)\Sigma_\parallel^{-1}r\;,\quad \gamma = -\frac{1}{2} \big(Q-\bar\Psi\big)^2 \Sigma_\parallel^{-1} \nonumber \;.
\end{gather}
In Fig.~\ref{fig:disppeakcontrib}, we show the explicit function $F(Q|r)$,
which is indeed of Gaussian form at a shifted position. Its width is of order the Lagrangian displacement dispersion, i.e. $\sim 10\hMpc$. After integrating over this function, we thus get to the correlation function of peaks in Eulerian space.
In the left-hand panel of Fig.~\ref{fig:disppeak}, we compare the resulting full correlation function of the Zeldovich displaced peaks as given by Eq.~(\ref{eq:lagconv-2}) to the evolved peak correlation function at linear order, which for the auto-correlation of the sample is given by \cite{Desjacques:2008jj}
\be
\label{eq:evolvedpkbias}
\xi(r)\approx (b_{10}+D_+)^2\xi_{0,0}(r)+2(b_{10}+D_+)(b_{01}-D_+ R_v^2)\xi_{2,0}(r)+(b_{01}-D_+ R_v^2)^2\xi_{4,0}(r)\, .
\ee
The linear bias part $(b_{10}+1)^2\xi_{0,0}$ is shown as the horizontal coloured lines on the right-hand panel of Fig.~\ref{fig:disppeak} where the ratio w.r.t.~the linear dark matter correlation is shown. We clearly see that both the haloes in the simulations and the peak predictions fall significantly below this linear bias prediction for scales between $20\hMpc$ and the BAO scale and are consistent one with the other. Below that scale, the peak prediction captures well the behaviour of the first stages of structure formation beyond linear theory but fails to capture the right amplitude of the bump and size of the exclusion zone towards lower redshifts (although the qualitative shape is similar).
Note that on the left-hand panel of Fig.~\ref{fig:disppeak}, we also display the correlation function of the proto-haloes displaced by the mean Zeldovich displacement field of their particles. As expected the exclusion zone is more pronounced in this case and is filled by the subsequent highly non-linear evolution.
The observed disagreement between the Zeldovich displaced proto-halo centers and Zeldovich displaced peaks is presumably due to the 20\% deviations between the peak model and the actual halo correlation and displacement discussed above in Figs.~\ref{fig:xibin4} and \ref{fig:vel}.

\section{Comparison to peaks in realizations}
\label{sec:realizations}
\begin{figure}[t]
\includegraphics[width=0.49\textwidth]{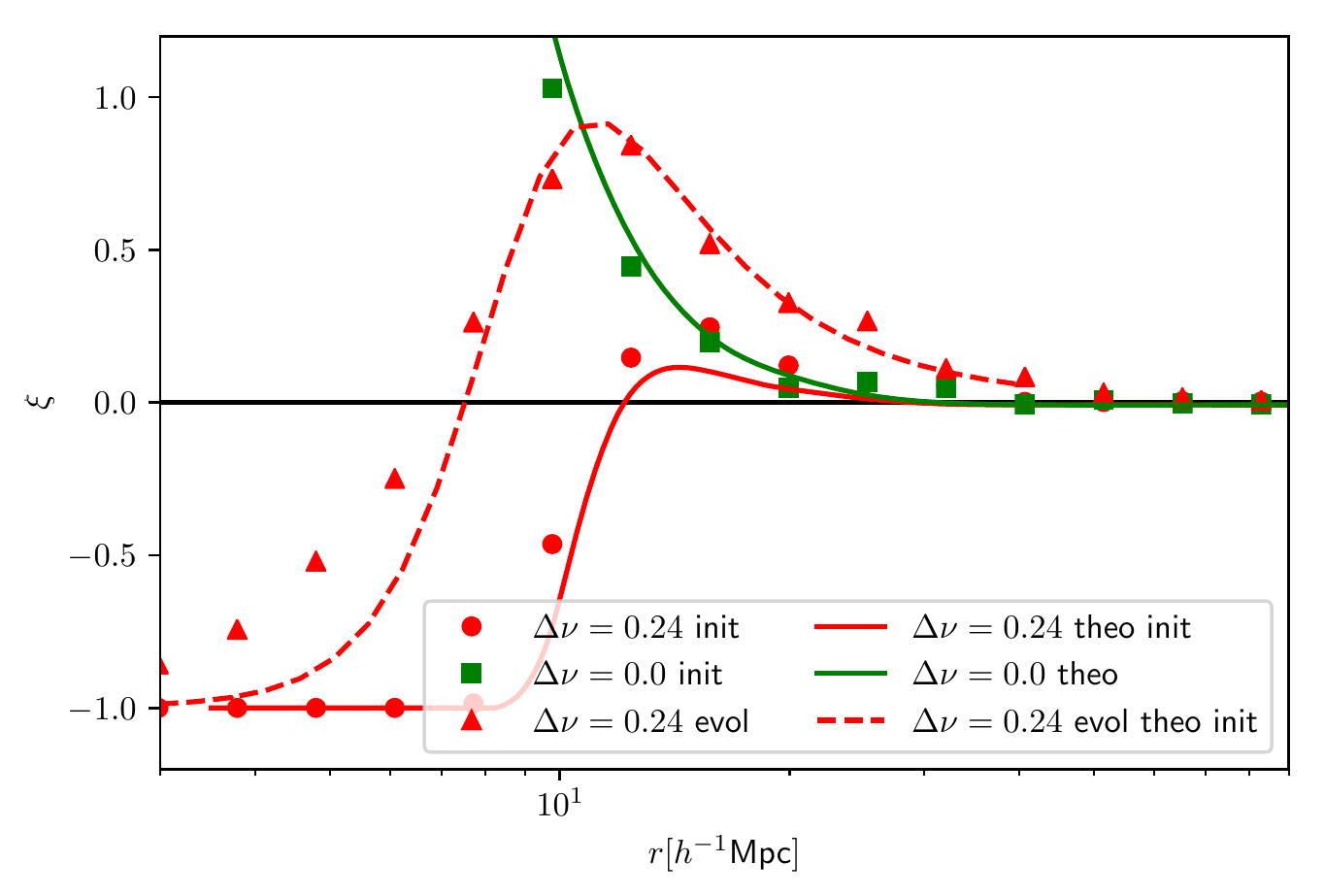}
\includegraphics[width=0.49\textwidth]{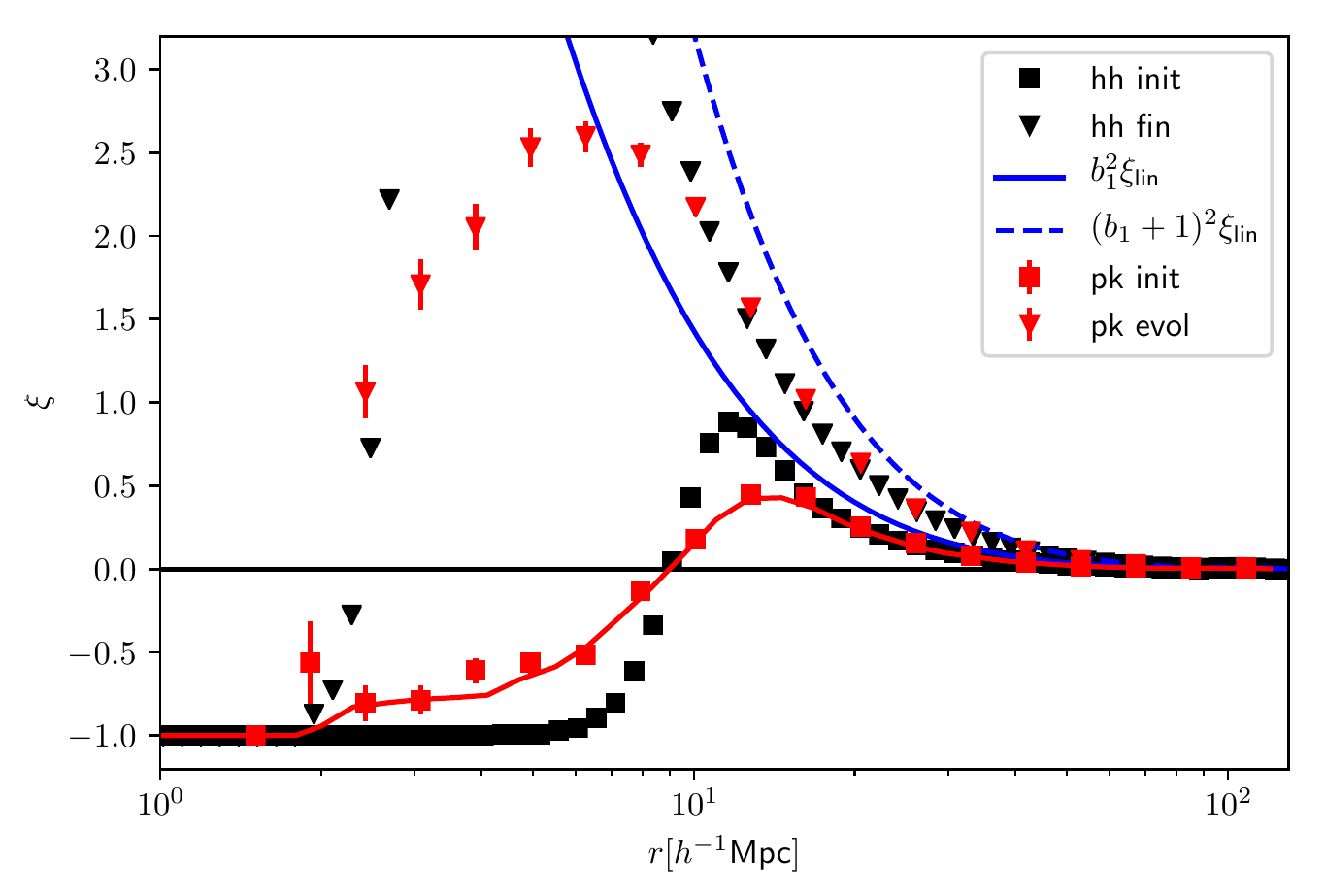}
\caption{\emph{Left panel: }Peak correlation functions averaged over 200 grid realizations for two distinct peak height bins of finite width $\delta\nu\approx 0.13$ with $\bar \nu=2.4$. The bin centers are separated by $\Delta \nu=0.24$ (red) and $\Delta \nu=0$ (green). For both cases we show the grid measurement and theory and find good agreement. We show the Zeldovich evolved peak correlation from the grid (red triangles) and theory (red dashed). \emph{Right panel: }Peaks selected from a Gaussian distribution of peak heights rather than peak height bins. Here we select a sample of maxima from a Gaussian distribution centered at $\bar \nu=2.4$ with standard deviation $\Delta \nu=0.24$ and show the auto-correlation of this sample (red squares) in comparison to the proto-haloes (black squares). Clearly the distribution of peak heights leads to effective exclusion, but the transition to the exclusion region is much smoother than what is seen for proto-haloes in the simulations. The evolved peak correlation function (red triangles) fails to reproduce the peak of the proto-halo correlation function (black triangles) outside the exclusion region.}
\label{fig:realpeakxi}
\end{figure}

To test and validate our results, we have implemented a grid-based peak finder that can be applied to realizations of cosmological density fields. We run this peak identification algorithm on the initial Gaussian density distribution of the simulation volume described above or alternative realizations in smaller volumes.
To select candidate grid cells for maxima, we first demand the density field at a grid vertex to be concave, i.e. the Hessian to be negative definite. We will assume that the Hessian be constant in the vicinity of the grid point $H_{ij}(\vec x_0)\approx H_{ij}(\vec x)$. By Sylvester criterion, the negative definiteness can be ensured by requiring the $k=1,2,3$ upper left minors $M_k$ of the Hessian to satisfy $(-1)^k M_k>0$. As a next step we check whether a maximum can be found in a unit cell centered on the grid point under consideration. For this purpose we expand the gradient of the density field as
\be
\nabla_i \delta(\vec x)\approx\nabla_i \delta(\vec x_0)+H_{ij}(\vec x_0)(\vec x-\vec x_0)_j\, .
\ee
Therefore, on requiring $\nabla_i \delta(\vec x)=0$, the separation of the maximum candidate from the grid point is simply given by
\be
(\vec x-\vec x_0)_i=H_{ij}^{-1}(\vec x_0)\nabla_j \delta(\vec x_0)\,.
\ee
We will associate a maximum to grid point $\vec x_0$ if $\max_i |(\vec x-\vec x_0)_i|<L/2 N_\text{c}$.
This Newton-method approach might lead to the identification of several maxima within one grid cell. To avoid this case, we pick the maximum with the largest amplitude within the cell.
We emphasize here that removing multiple peaks from a single cell does not induce any exclusion. The cell sizes employed in this study are significantly smaller than the typical exclusion separations observed in our measurements and we have ensured convergence by runs with smaller grid sizes. 

In Fig.~\ref{fig:realpeakxi}, we show the result of the realization sample for the $R=4.3\hMpc$, $\bar \nu=2.4$ sample corresponding to simulation halo mass bin IV. We select peaks using a cubic grid with $N_\text{c}=512$ cells per dimension in a cubic box with side length $L=400 \hMpc$. The small volume helps with resolution on small scales, but requires us to average over $200$ realizations to reduce the error bars. For the numerical implementation $\Delta \nu=0$ case, we select a narrow bin of peak heights $\nu \in [2.35,2.45]$. For the $\Delta \nu\neq 0$ case, we cross-correlate peaks from two samples with $\nu_A\in[2.20,2.33]$ and $\nu_B\in[2.47,2.60]$.

As we can see, both in the initial conditions and in the evolved field, the grid results are in very good agreement with the numerical sampling presented in the main text of this paper. The initial conditions show a clear exclusion regime and the evolution both increases the clustering amplitude on intermediate scales and reduces the exclusion radius.

We also sample peak heights from a Gaussian distribution, aiming to reproduce the actual distribution of peak heights observed for the proto-haloes in the simulation (see Appendix \ref{app:scatter}). As shown in the right panel of Fig.~\ref{fig:realpeakxi}, this leads to a significantly smoother transition to the exclusion regime.


\section{Conclusions}
\label{sec:conclusion}

This study presents the first implementation of the \emph{non-perturbative} correlation function of peaks of the linear, three-dimensional density field (Lagrangian space) and its perturbative evolution to the halo formation epoch (Eulerian space). 
Unlike other numerical implementations of perturbative bias expansions, the approach considered here deals with discrete tracers from the onset. Our results can be summarized as follows:
\begin{itemize}
\item At fixed smoothing scale, unequal height peaks exhibit exclusion while equal height peaks do not. This arises from the fact that, on a given smoothing scale, a single local density maximum can be split into two nearby peaks at no cost. Using this approach we can reproduce the clustering of proto-haloes observed in $N$-body simulations.
\item The above behaviour can be analytically and therefore quantitatively understood for signed critical points.
\item Imposing an upper bound (lower bound on the absolute value as peaks have negative curvatures) on the three negative eigenvalues of the Hessian can generate exclusion even for equal height peaks.
\item Peak velocity statistics deviate from the underlying matter velocity statistics significantly. These deviations are in accordance with what is seen for proto-haloes in $N$-body simulations.
\item We derive a closed form expression for the {\it non-perturbative} clustering of Zeldovich displaced peaks and study its behaviour down to the smallest scales (the exclusion region). The non-perturbative, evolved peak clustering reproduces the halo-clustering down to separations of $10-20 \hMpc$. Like for haloes, the evolved peaks exclusion region shrinks with time. However, our peak-based prediction fails to reproduce the detailed shape of the transition from exclusion to mildly non-linear regime as measured for haloes.
\item While in this study we apply a Gaussian filter, there is evidence for a mixed Gaussian plus top-hat filtering being in better agreement with simulations \cite{Chan:2015zjt}. Improvement could also arise from \cite{Biagetti:2013hfa} taking into account the upcrossing constraint which, for the Gaussian smoothing employed here, amounts to a simple multiplicative weight \citep{Appel:1990mfb,paranjape/sheth:2012}.
\end{itemize}

As we have shown in this study, the statistics of halo displacements do differ from the matter displacements in a way that is captured by the peak model. Halo displacements are at the core of reconstruction techniques that aim to undo the effect of long-wavelength motions. In these methods the halo displacement is estimated from the smoothed halo correlation function, ignoring scale-dependent density and velocity bias. We expect that accounting for these distinct scale dependencies will improve the performance of reconstruction algorithms for future surveys.

Furthermore, in the halo model, galaxy correlation functions are calculated by convolving the distribution of halo centers with the corresponding matter (galaxy) profile. In its standard implementation, the halo model relies on a linear bias model for the correlation of halo centers and leads to an unphysical constant contribution for low wavenumbers.
As pointed out by \cite{Smith:2010fh,Hamaus:2010im,Schmidt:2015gwz}, this is related to the halo stochasticity covariance.
Finally, we also anticipate that a non-perturbative description of the two-halo contribution along the lines considered here could be helpful toward a more accurate description of the transition region between the two-halo and the one-halo terms \cite{Jose:2017jod,hadzhiyska2020galaxyhalo}. For specific galaxy populations, the exclusion region may even be visible in the real space correlation function (see e.g. Fig.~9 of \cite{Desjacques:2003ab}). In such cases, a perturbative description of the two-halo term would not be accurate enough. 

\begin{acknowledgements}
TB is supported by a Stephen Hawking Advanced Fellowship at the Center for Theoretical Cosmology, University of Cambridge. SC's work is partially supported by the SPHERES grant ANR-18-CE31-0009 and by Fondation MERAC. 
VD acknowledges support by the Israel Science Foundation (grants no. 1395/16 and 2562/20). 
CP is supported by the Segal grant ANR-19-CE31-0017 (secular-evolution.org) of the French {\sl Agence Nationale de la Recherche}.
TB would like to thank Kacper Kornet for excellent computing support. This work has also made use of the Horizon Cluster hosted by Institut d'Astrophysique de Paris. We thank Stephane Rouberol for running this cluster smoothly for us. 
SC thanks Takahiko Matsubara for fruitful discussions. CP thanks Simon Prunet for early discussions and Dmitry Pogosyan, Junsup Shim and Corentin Cadiou for feedback.
\end{acknowledgements}

\bibliography{3dpeak}

\appendix
\section{Signed critical points}\label{app:critpoints}
In this Appendix, we discuss the technical details of the derivation of the closed-form expression for the correlation function of signed critical points discussed in Sec.~\ref{sec:signedcritpoints}.
We first split the state vector into $\vec m^\text{T}=(\sigma_0 \nu_1,\sigma_1 \vec \eta_1,\sigma_0 \nu_2,\sigma_1 \vec \eta_2)$ and $\vec n^\text{T}=(\sigma_2 \vec \zeta_1,\sigma_2 \vec \zeta_2)$ so that
\be
\begin{split}
1+\xi_\text{crit}=&\frac{1}{{\bar n}^2_\text{crit}}\int \derd^8\vec m\; w_m(\vec m)\int \derd^{12}\vec n \det[H_{1,ij}]\det[H_{2,ij}]\exp\left[-\frac{1}{2}\vec m^\text{T}\scalp \vec \Omega_{\vec m,\vec m} \scalp\vec m-\vec m^\text{T} \scalp\vec \Omega_{\vec m,\vec n} \scalp\vec n-\frac{1}{2} \vec n\scalp\vec \Omega_{\vec n,\vec n} \scalp\vec n\right]\ .\\
=&\frac{1}{{\bar n}^2_\text{crit}}\int \derd^8\vec m\; w_m(\vec m)\det[\partial_{\beta_{P_{ij}}}]\det[\partial_{\beta_{O_{ij}}}] \int \derd^{12}\vec n \exp\left[-\frac{1}{2}\vec m^\text{T} \scalp\vec \Omega_{\vec m,\vec m} \scalp\vec m-\vec\beta^\text{T} \scalp\vec n-\frac{1}{2} \vec n^\text{T}\scalp\vec \Omega_{\vec n,\vec n} \scalp\vec n\right].
\end{split}
\ee
Here we have written the components of the determinant prefactors as derivative operators with respect to the components of $\vec \beta^\text{T}=\vec m^\text{T}\vec \Omega_{\vec m,\vec n}$ with
\begin{align}
P=\begin{pmatrix}
1 & 4& 5\\
4& 2 & 6\\
5 & 6& 3
\end{pmatrix}\, ,
&&
O=\begin{pmatrix}
7 & 10& 11\\
10& 8 & 12\\
11 & 12& 9
\end{pmatrix}\, ,
\end{align}
and we have defined $w_m(\vec m)=\delta^\text{(D)}[\sigma_1 \vec \eta_1]\delta^\text{(D)}[\sigma_1 \vec \eta_2].$
We can now perform the Gaussian integral over $\vec n$ and obtain
\be
1+\xi_\text{crit}=\frac{1}{{\bar n}^2_\text{crit}}\int \derd^8\vec m\;\det[\partial_{\beta_{P_{ij}}}]\det[\partial_{\beta_{O_{ij}}}]\exp\left[-\frac{1}{2}\vec m^\text{T} \vec \Omega_{\vec m,\vec m} \vec m+\frac{1}{2}\vec\beta^\text{T} \vec \Omega_{\vec n,\vec n}^{-1} \vec \beta\right].
\ee
Taking the $\beta$ derivatives and combining terms in the exponential\footnote{The inversion of block matrices yields for the relation of the blocks of the covariance and precision matrix that $\vec C_{\vec m,\vec m}^{-1}=\vec \Omega_{\vec m,\vec m}-\vec \Omega_{\vec m,\vec n}^\text{T}\vec\Omega_{\vec n,\vec n}^{-1}\vec \Omega_{\vec m,\vec n}$.}
yield
\be
1+\xi_\text{crit}=\frac{1}{{\bar n}^2_\text{crit}}\int \derd^8\vec m\; w_m(\vec m) \mathcal{D}\exp\left[-\frac{1}{2}\vec m^\text{T} \scalp\vec C_{\vec m,\vec m}^{-1}\scalp\vec m\right].
\ee
The $m$-integration collapses due to the weight function $w_m(\vec m)$, which sets the gradients to zero and the peak heights to a specific value.
Taking the derivatives, the prefactor evaluates to
\be
\begin{split}
\mathcal{D}\sim& \epsilon_{i_1,i_2,i_3}\epsilon_{j_1,j_2,j_3}\Biggl\{\prod_{i=1}^6 \vec m \vec C_{\vec m,\vec m}^{-1}\vec C_{\vec m,\vec n} \vec p^{(\kappa_i)}+
\sum_{\kappa}\Bigl[\vec \Omega^{-1}_{\vec n,\vec n;\kappa_5,\kappa_6} \prod_{i=1}^4 \vec m \vec C_{\vec m,\vec m}^{-1}\vec C_{\vec m,\vec n} \vec p^{(\kappa_i)} \\
&
+\vec m \vec C_{\vec m,\vec m}^{-1}\vec C_{\vec m,\vec n} \vec p^{(\kappa_5)}\ \vec  m \vec C_{\vec m,\vec m}^{-1}\vec C_{\vec m,\vec n} \vec p^{(\kappa_6)}  \prod_{i,j}^4 \vec \Omega^{-1}_{\vec n,\vec n;\kappa_i,\kappa_j}+\prod \vec \Omega^{-1}_{\vec n,\vec n;\kappa_i,\kappa_j}\Bigr]\Biggr\}\; ,
\end{split}
\ee
where $\kappa$ is a permutation of the derivative indices $(P_{1,i_1},P_{2,i_2},P_{3,i_3},O_{1,j_1},O_{2,j_2},O_{3,j_3})$ and where $p^{(i)}_j=\delta^\text{(K)}_{ij}$.
Eventually, we get
\be
1+\xi_\text{crit}\approx  A(r)e^{B(r)},
\ee
with
\be
A(r)=\frac{\mathcal{D}}{\bar n_\text{crit}^2}\, ,
\ee
and
 \be
B(r)=-\frac12\vec m^\text{T} \vec C_{\vec m,\vec m}^{-1}\vec m=\frac{\nu ^2}{4}+\frac{7 \Delta \nu ^2}{80}-\frac{96 \Delta \nu
   ^2}{\tilde r^6}+\frac{12 \Delta \nu
   ^2}{\tilde r^4}-\frac{6 \Delta \nu ^2}{5 \tilde r^2}-\frac{81 \Delta \nu ^2
   \tilde r^2}{22400}-\frac{\nu ^2 \tilde r^2}{32}+\frac{11 \Delta \nu ^2 \tilde r^4}{179200}+\frac{\nu ^2 \tilde r^4}{768}\; .
\ee
This exponential suppresses $1+\xi$ for non-vanishing peak height differences $\Delta \nu \neq 0$.

\section{Curvature Cutoff}
\begin{figure}[t]
\centering
\includegraphics[width=0.49\textwidth]{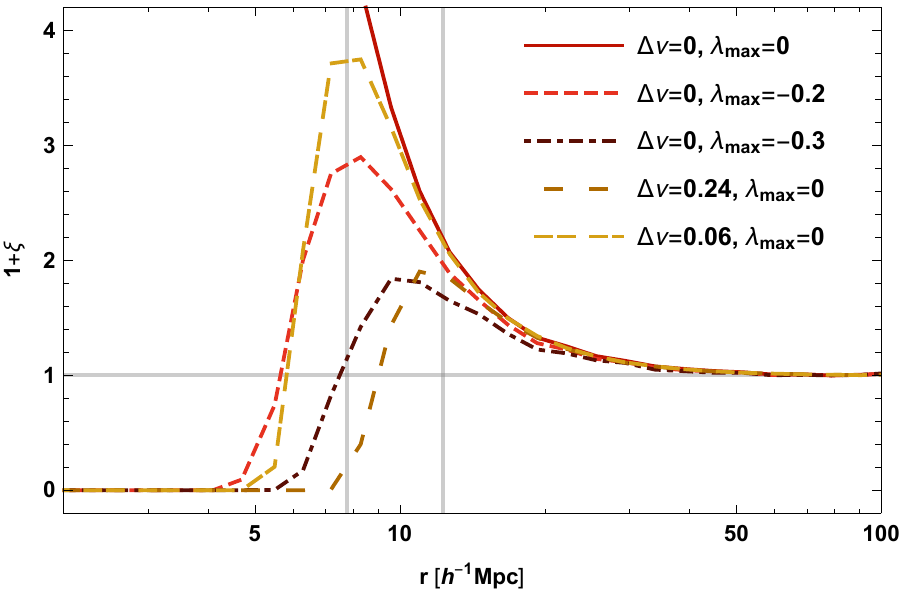}
\caption{Effect of finite peak height separation $\Delta \nu$ and peak curvature cutoff $\lambda_\text{max}$ on exclusion in the correlation function. Imposing an upper limit on the largest eigenvalue (lower limit on the lowest magnitude eigenvalues) of the Hessian leads to exclusion effects similar to the finite peak height difference.
}
\label{fig:curvcut}
\end{figure}

While we had already explored the effect of non-vanishing peak height difference in BCDP, we hadn't explored the effect of a cutoff in peak curvature on exclusion. In practice we implement this cutoff as an upper bound $\lambda_\text{max}$ on the largest (lowest magnitude) eigenvalue of the ordered set $\lambda_3<\lambda_2<\lambda_1<\lambda_\text{max}<0$. As we show in Fig.~\ref{fig:curvcut}, the curvature cutoff does indeed lead to small-scale exclusion with $R_\text{excl}\approx 6 R\lambda_\text{cut}^{2/3}$. When measuring the eigenvalues of the Hessian at the proto-halo positions, we do not find such a cutoff. This might be due to the fact that our single Gaussian filter is too simplistic. Furthermore, \cite{Ludlow:2011tpf} have found that a fraction of the haloes do actually form at the saddle point between two peaks.

\section{Abundance of peaks and critical points and bias parameters}
\label{sec:bias}
The eigenvalues $\lambda_i$ of the Hessian $H_{ij}$ can be rewritten as
\begin{align}
x&=\lambda_1+\lambda_2+\lambda_3\,, \\
y&=\lambda_1-\lambda_3\,, \\
z&=\lambda_1-2\lambda_2+\lambda_3\,.
\end{align}
When calculating the abundance, we can integrate out $y$ and $z$ analytically yielding
\be
\bar n_\text{pk}=\frac{1}{(2\pi)^2R_\star^3}\int \derd \nu w(\nu) e^{-\frac{\nu^2}{2}}G_0(\nu,\gamma\nu)\, ,
\ee
where
\be
G_i(\nu,\gamma \nu)=\int \derd x \ x^if(x)\frac{1}{\sqrt{2\pi(1-\gamma^2)}}e^{-\frac{(x-\gamma \nu)^2}{2(1-\gamma^2)}}\,,
\ee
with 
\be
f(x)=\sqrt{\frac{2}{5\pi}}\frac{1}{20}\left[\left(10x^2-32\right)e^{\frac{-5x^2}{2}} + (155x^2+32)e^{\frac{-5x^2}{8}}\right]
+\frac12 x(x^2-3)
            \left[\text{erf}\left(\sqrt{\frac{5}{2}}\frac{x}{2}\right) +\text{erf}\left(\sqrt{\frac{5}{2}}x\right)\right].
\ee

The bias parameters are then commonly defined as
\be
b_{ij}(\nu)=\frac{1}{\bar n_\text{pk}}\frac{1}{(2\pi)^2R_\star^3}\int \derd x \tilde b_{ij}(\nu,x) f(x)e^{-\frac{\nu^2}{2}}\frac{1}{\sqrt{2\pi(1-\gamma^2)}}e^{-\frac{(x-\gamma \nu)^2}{2(1-\gamma^2)}}\,,
\ee
where the explicit expression for the coefficients up to second order are given by
\be
\begin{split}
\tilde b_{10}(\nu,x) =&\frac{1}{\sigma_0}\frac{\nu-\gamma x}{1-\gamma^2},\ \ \Rightarrow \ \ b_{10}(\nu) =\frac{1}{\sigma_0}\frac{\nu-\gamma \bar x}{1-\gamma^2},\\
\tilde  b_{01}(\nu,x) =&\frac{1}{\sigma_2}\frac{x-\gamma \nu}{1-\gamma^2}, \ \ \Rightarrow \ \ b_{01}(\nu) =\frac{1}{\sigma_2}\frac{\bar x-\gamma \nu}{1-\gamma^2},\\
\end{split}
\ee
where $\bar x=G_1(\nu)/G_0(\nu)$.

\section{Scatter in the Simulations}\label{app:scatter}
\begin{figure}
\centering
\includegraphics[width=0.49\textwidth]{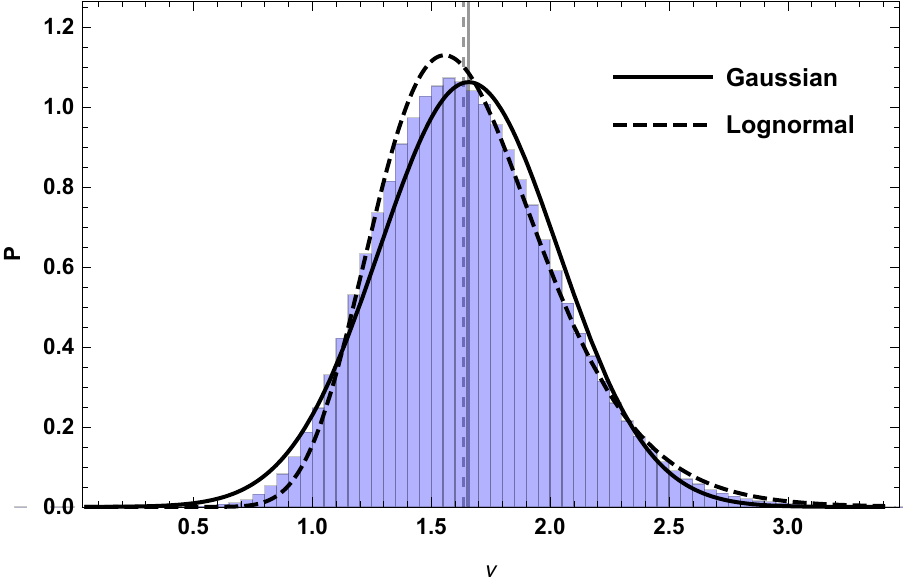}
\includegraphics[width=0.49\textwidth]{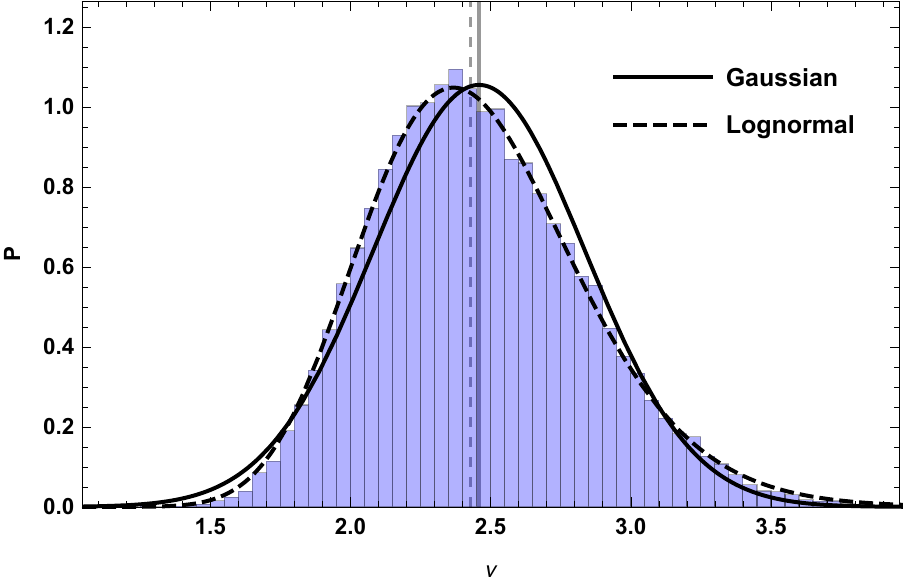}
\caption{PDFs of the density at protohalo position for bins II and IV. We overplot a Gaussian distribution (solid) and log-normal distribution (dashed), the latter providing a better discription of the measurements. The vertical dashed line gives the mean of the measured distribution and the vertical dashed line gives the median.}
\label{fig:hist}
\end{figure}

In Fig.~\ref{fig:hist} we show the distribution of smoothed densities at the proto-halo position normalized by the standard deviation of the smoothed density field $\sigma_0$ for bins II and IV. The smoothing scales are given in Tab.~\ref{tab:massbins} and is motivated by fits to the cross-power spectrum between proto-haloes and the underlying Gaussian density field \cite{Baldauf:2014fza}.
The distribution of measured densities at the proto-halo position is well described by a log-normal distribution but not too far off from a Gaussian distribution. 
We have checked that the difference between both is minor.

\end{document}